\documentclass[%
 aip,
 amsmath,amssymb,
 reprint,%
]{revtex4-1}
\usepackage{mathrsfs,bm}
\usepackage{longtable}
\usepackage{txfonts}
\usepackage{indentfirst}
\usepackage{graphicx,color,dcolumn,booktabs}
\usepackage{multirow}
\usepackage{indentfirst}
\usepackage{multirow}
\usepackage{epsfig}
\usepackage{graphicx,color,dcolumn}
\usepackage{epstopdf}
\usepackage{amsmath}
\usepackage{multirow}
\usepackage{diagbox}
\usepackage{changes}
\usepackage{threeparttable}
\usepackage{enumitem,overpic,chemarr}

\usepackage{color}
\usepackage{amssymb}
\definecolor{cover}{rgb}{0.77,0.87,0.88}
\definecolor{blueone}{rgb}{0.1,0.1,.7}
\definecolor{citec}{rgb}{0.14,0.47,0.09}
\definecolor{two}{rgb}{0.0,0.5,0.}
\definecolor{three}{rgb}{.5,.1,0.15}
\usepackage[bookmarks=true,bookmarksopen=false,plainpages=false,breaklinks=true,
   bookmarksnumbered=true,hypertexnames=false,
   filecolor=blue,urlcolor=three,menucolor=three,
   linkcolor=three,citecolor=blueone, colorlinks,
   anchorcolor=blue,runcolor=pink,frenchlinks=red
   pdfstartview=FitH,pdftitle=title,%
   pdfauthor=author]{hyperref}

\usepackage{relsize}
\usepackage{xspace}
\def\babar{\mbox{\slshape B\kern-0.1em{\smaller A}\kern-0.1em
    B\kern-0.1em{\smaller A\kern-0.2em R}}}

\allowdisplaybreaks[4]
\usepackage{color}
\usepackage{graphicx}% Include figure files
\usepackage{dcolumn}% Align table columns on decimal point
\usepackage{bm}% bold math
\usepackage[utf8]{inputenc}
\usepackage[T1]{fontenc}
\usepackage{mathptmx}
\usepackage{etoolbox}

\makeatletter
\def\@email#1#2{%
 \endgroup
 \patchcmd{\titleblock@produce}
  {\frontmatter@RRAPformat}
  {\frontmatter@RRAPformat{\produce@RRAP{*#1\href{mailto:#2}{#2}}}\frontmatter@RRAPformat}
  {}{}
}%
\makeatother
\begin{document}

\preprint{AIP/123-QED}

\title[]{Formation and Regulation of Calcium Sparks on a Nonlinear Spatial Network of
Ryanodine Receptors}
% Force line breaks with \\
\author{Tian-Tian Li}
 \altaffiliation[]{These authors have contributed equally to this work.}
\affiliation{School of Physics and Technology, Nanjing Normal University,
  Nanjing 210097, China} %Lines break automatically or can be forced with \\
\author{Zhong-Xue Gao}%
 \altaffiliation[]{These authors have contributed equally to this work.}
\affiliation{School of Physics and Technology, Nanjing Normal University,
  Nanjing 210097, China}%
\author{Zuo-Ming Ding}%
\affiliation{School of Physics and Technology, Nanjing Normal University,
  Nanjing 210097, China}%
\author{Han-Yu Jiang}
\email{junhe@njnu.edu.cn }
\email{jianghy@njnu.edu.cn}
\altaffiliation[]{Corresponding author}
\affiliation{School of Physics and Technology, Nanjing Normal University,
  Nanjing 210097, China} %
\author{Jun He}

\altaffiliation[]{Corresponding author}
\affiliation{School of Physics and Technology, Nanjing Normal University,
  Nanjing 210097, China} %
\date{\today}% It is always \today, today,
             %  but any date may be explicitly specified

\begin{abstract} Accurate regulation of calcium release is essential for
cellular signaling, with the spatial distribution of ryanodine receptors (RyRs)
playing a critical role. In this study, we present a nonlinear spatial network
model that simulates RyR spatial organization to investigate calcium release
dynamics by integrating RyR behavior, calcium buffering, and calsequestrin (CSQ)
regulation.  The model successfully reproduces calcium sparks, shedding light on
their initiation, duration, and termination mechanisms under clamped calcium
conditions. Our simulations demonstrate that RyR clusters act as on-off switches
for calcium release, producing short-lived calcium quarks and longer-lasting
calcium sparks based on distinct activation patterns. Spark termination is
governed by calcium gradients and stochastic RyR dynamics, with CSQ facilitating
RyR closure and spark termination.  We also uncover the dual role of CSQ as both
a calcium buffer and a regulator of RyRs. Elevated CSQ levels prolong calcium
release due to buffering effects, while CSQ-RyR interactions induce excessive
refractoriness, a phenomenon linked to pathological conditions such as
ventricular arrhythmias. Dysregulated CSQ function disrupts the on-off switching
behavior of RyRs, impairing calcium release dynamics.  These findings provide
new insights into RyR-mediated calcium signaling, highlighting CSQ's pivotal
role in maintaining calcium homeostasis and its implications for pathological
conditions. This work advances the understanding of calcium spark regulation and
underscores its significance for cardiomyocyte function. \end{abstract}
\maketitle

\begin{quotation} Calcium release events, particularly calcium sparks mediated
by ryanodine receptors, are vital for cellular signaling, underpinning processes
like muscle contraction and neuronal communication. Advances in super-resolution
imaging have revealed the intricate spatial organization of RyRs, highlighting
their crucial role in calcium dynamics. This study introduces a nonlinear
spatial network model to simulate calcium release, incorporating the effects of
RyR cluster distribution and the regulatory role of calsequestrin. By
replicating spontaneous and evoked calcium sparks, the model elucidates how RyR
transition rates, calcium concentrations, and CSQ buffering influence spark
duration and refractoriness. These insights shed light on calcium signaling
regulation and its implications for diseases like ventricular tachycardia.
\end{quotation}

\maketitle

\section{Introduction}\label{sec1}

Calcium release events (CREs), particularly calcium sparks, mediated by
RyRs on the junctional sarcoplasmic reticulum (JSR), are
crucial for sustaining cellular functions through complex calcium signaling
pathways, especially in cardiac and skeletal muscle cells.~\cite{Wehrens2005,Cheng2008,Terentyev2008,Endo2009,Bers2013,Berridge2016,Eisner2017}
Under physiological conditions, RyRs exhibit a low probability of opening,
whereas pathological states can lead to malfunction-induced activation. This
activation of RyRs has a minimal likelihood of initiating calcium release within
entire RyR clusters, acting as an on-off switch.~\cite{Asfaw2013,Song2016} When
this occurs, spontaneous calcium sparks are governed by calcium-induced calcium
release (CICR) mechanisms.~\cite{Cheng1993} Similarly, calcium release of
 L-type calcium channels (LCCs) can also evoke calcium sparks, exhibiting
experimental behavior comparable to spontaneous sparks.~\cite{Wang2001,Lopez-Lopez1995} 
These localized calcium sparks
play a pivotal role in various physiological processes, including
excitation-contraction coupling in cardiac and muscle cells, as well as synaptic
transmission in neurons.~\cite{Eisner2017,Bers2002,Ross2012,Feske2007} Moreover,
calcium sparks form the fundamental building blocks of larger-scale calcium
waves. Thus, calcium sparks occupy a central position within the hierarchical
framework governing calcium dynamics within cells, necessitating a profound
understanding of their formation and regulation to elucidate the mechanisms of
calcium release and their implications in disease pathogenesis.

Early investigations into calcium release often lacked explicit representations
of cluster structures and relied on mean-field assumptions, which overlooked
CICR interactions among RyRs within a cluster.~\cite{Sobie2002, Williams2011}
This mean-field approach failed to capture the localized dynamics of calcium
release, as evidenced by numerous studies that highlighted its
limitations.~\cite{Sobie2002, Williams2011, Sneyd2017, Keizer1997, Cohen2019,
Iaparov2019, Iaparov2021, Cannell2013}  More recent research has addressed
these concerns by incorporating explicit CICR mechanisms within the cluster
structure.~\cite{Cannell2013, Walker2014, Walker2015, Iaparov2021, Kolstad}
Similarly, the spatial distribution of inositol 1,4,5-trisphosphate receptors
(IP3R), another type of calcium channel, has also been shown to be crucial
for understanding their functional properties.~\cite{Berridge2016, Shuai2002,
Smith2009, Rudiger2014,Scott2021}  Advanced imaging techniques, such as electron
tomography and single-molecule localization microscopy, have revealed that
RyR clusters exhibit irregular shapes, varying sizes, and a stochastic
distribution of RyRs, contrary to earlier assumptions of densely packed,
lattice-like configurations.~\cite{Franzini-Armstrong1999, Hurley2023,
Baddeley2009, Jayasinghe2018, Shen2019} This intricate organization further
underscores the importance of incorporating cluster-specific mechanisms into
calcium release models.

Pioneering efforts in the literature have studied calcium signal transduction on
networks of calcium channels, incorporating the spatial distribution
characteristics of RyRs. Notably,  Walker et al.~\cite{Walker2014, Walker2015}
introduced an adjacency matrix for modeling calcium release in the heart,
simulating lattice-like configurations of RyRs within clusters. Further
approaches, as explored in the literature~\cite{Iaparov2019, Iaparov2021,
Hernandez-Hernandez2017} and in our previous studies,~\cite{Jiang2021, Gao2023}
conceptualized RyRs as nodes of a network interconnected by the diffusion of
calcium ions. Given these advancements, it is intriguing to study calcium
release with new experimental information on RyR spatial distribution, which has
not been previously considered in investigations of calcium signaling.

Calcium sparks occur in the dyad between the JSR and the transverse tubule.
These sparks are influenced not only by the spatial distribution of RyRs but
also by calcium concentrations and buffers, particularly CSQ, which plays a
critical role in calcium release.~\cite{Eisner2023} Calcium sparks in transgenic mice with overexpressed
or underexpressed CSQ exhibit longer or shorter duration of calcium sparks,
respectively, highlighting the crucial role of CSQ in modulating RyR activity.~\cite{Terentyev2003}
The authors suggested that the predominant effects of CSQ on JSR calcium release
are due to its buffering actions inside the JSR.  CSQ regulates calcium sparks
not only by acting as a buffer but also by influencing RyR
activity.~\cite{Damiani1994, Chen2013, Sun2021} Lipid bilayer experiments have
shown that RyR channel activity is modulated by the interaction of CSQ with the
auxiliary proteins triadin and junctin.~\cite{Zhang1997, Gyoke2004, Beard2002}
Specifically, CSQ dissociates from RyR under high JSR calcium concentration, enhancing RyR
sensitivity to ${\rm Ca}^{2+}$ in the dyadic cleft. Conversely, low JSR calcium concentration causes CSQ
binding to RyR, diminishing sensitivity.  Raising the luminal
calcium concentration from 20~$\mu$M to 5 mM, which binds to and polymerizes
CSQ, leads to an enhanced open probability of the channels.~\cite{Gyorke2004} 

These findings highlight the intricate regulatory roles of CSQ in calcium
signaling dynamics, with increasing attention being paid to their relevance to
diseases such as catecholaminergic polymorphic ventricular tachycardia (CPVT) in
recent years.~\cite{Liu2009, Sibbles2022, Sun2021}  However, there is still a
need for more simulation and theoretical analysis to fully understand this
phenomenon.  Restrepo et al. simulated the effect of CSQ by using
CSQ-bound and CSQ-unbound states,~\cite{Restrepo2008} revealing its contribution to the excessive
refractoriness of calcium sparks.~\cite{Brochet2005} However, their simulation,
which considered more calcium cycling mechanisms, and 
a realistic number of dyads, each with a realistic number of RyR channels, makes it
computationally difficult to incorporate the spatial distribution of RyRs or the
explicit CICR mechanism.~\cite{Restrepo2008} 

Incorporating spatial distributions into the RyR transition model offers a novel
pathway for calcium signaling research, particularly through the application of
recent experimental data,~\cite{Hurley2023, Baddeley2009,
Jayasinghe2018, Shen2019} which remain underexplored in CRE studies. Our goal is to simulate CREs to assess whether calcium
sparks with experimentally observed properties, specifically their on-off
switching behavior, can be accurately reproduced. By utilizing this model, we
aim to explore the regulation of calcium sparks through clamped calcium
concentrations in the myoplasm and network sarcoplasmic reticulum (NSR).
Additionally, we plan to investigate the explicit influence of calsequestrin on
RyR clusters,\cite{Terentyev2003, Damiani1994, Chen2013, Sun2021,Brochet2005}
especially by incorporating spatial RyR distributions that are underrepresented
in existing literature. This approach is expected to provide deeper insights
into the buffering mechanisms of calsequestrin and its impact on RyR function,
thereby offering multiple directions for follow-up studies in cellular signaling
regulation.

In this study, we introduce a spatial network model in Section~\ref{Mechanism}
to incorporate the spatial distribution of RyRs.  We account for CSQ
 binding and unbinding states to capture its effects on RyR activity.
Additionally, RyR transition rates are determined by fitting experimental data.
The calcium exchange and calcium buffers are also considered to establish a
mechanism for simulating calcium release through RyRs on JSR. In
Section~\ref{results}, the simulation of CREs aims to replicate its on-off
switch characteristic, specifically focusing on the formation of calcium sparks
with spontaneous and evoked activations. In Section~\ref{Regulation}, we discuss the regulation 
of calcium sprak by opening rate
and clamped calcium concentrations in the myoplasm and NSR concerning the
duration and open RyRs of calcium sparks. Moreover, we investigate the
regulation of calcium sparks by CSQ, which affects the opening possibilities of
RyRs and acts as a buffer, We use experimental data to elucidate these roles and
discuss potential malfunctions in CSQ regulation. The article concludes with
discussion and summary in the last section~\ref{Summary}.

\section{Model of Calcium Release on JSR}\label{Mechanism}

\begin{figure}[h!]
  \includegraphics[bb=120 250 490 500,scale=0.65,clip]{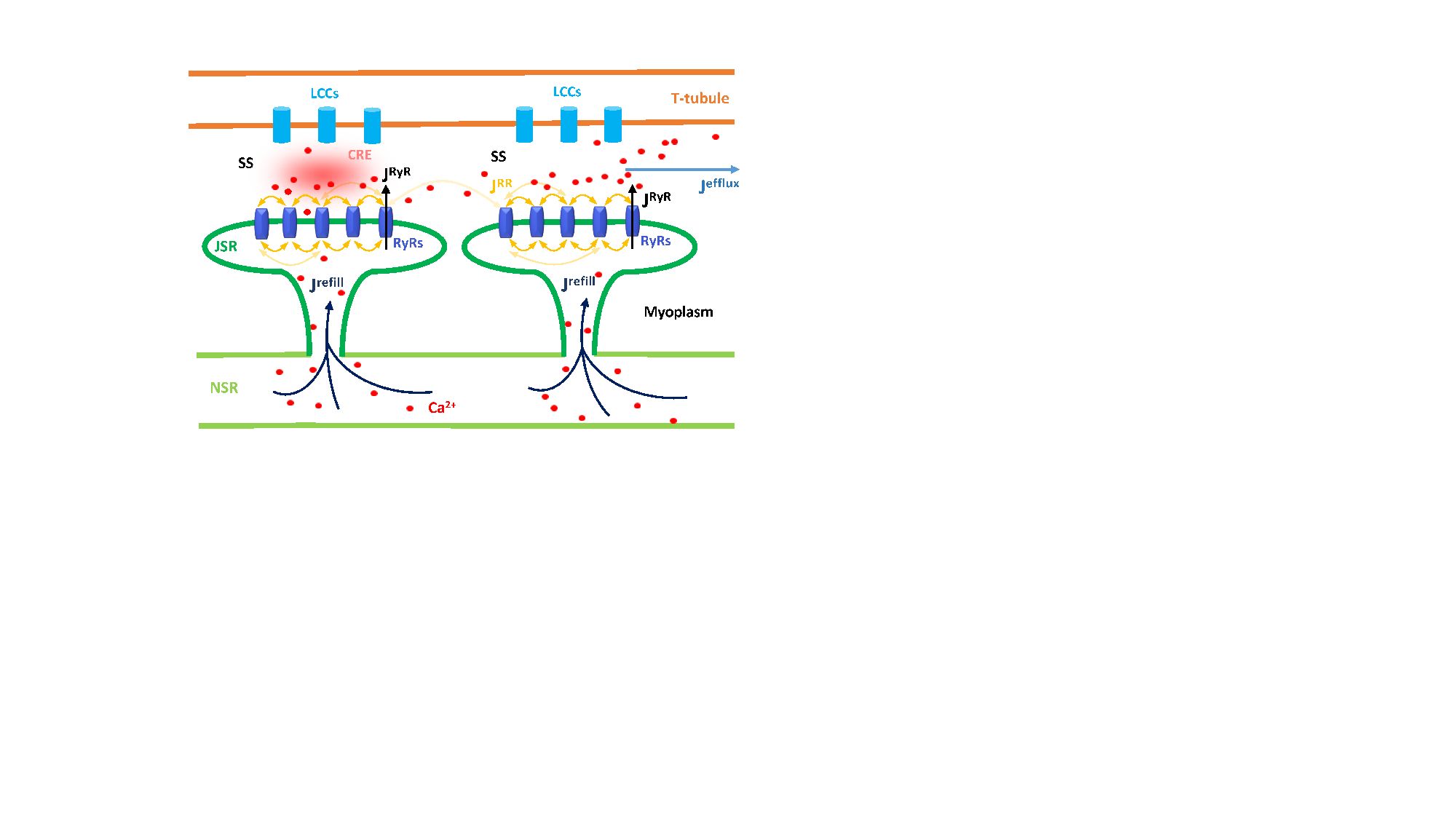}
  \caption{Schematic representation of the calcium release mechanism. The RyRs,
  located in the JSR, facilitate the flux of calcium ions ($J_{i}^{\rm RyR}$)
  from the JSR to the SS. Calcium ions also diffuse from other RyRs
  to a specific RyR $i$ in both the JSR and SS, denoted as a flux $J_{i}^{\rm
  RR}$. These fluxes collectively contribute to CREs
  occurring within the SS. Calcium ions in the SS can exit into the myoplasm
  through efflux, represented as $J_i^{\rm efflux}$. Simultaneously, calcium
  ions refill the JSR from the NSR, depicted as $J_i^{\rm refill}$. }
  \label{Fig: schematic}
\end{figure}

This study focuses on elucidating the complexities of CREs, particularly calcium
sparks, occurring on the JSR. The JSR, a
terminal cistern of the sarcoplasmic reticulum containing an array of RyRs, is
dispersed randomly around the transverse tubule, forming a subspace (SS) with an
average distance of approximately 15 nm as shown in Fig.~\ref{Fig: schematic}. The release of calcium ions from the
JSR into the SS is orchestrated by the controlled opening of RyRs. The state of
these receptors, whether open or closed, is contingent upon the local
concentrations of calcium ions, denoted as $[{\rm Ca}^{2+}]_i^{\rm SS}$ and
$[{\rm Ca}^{2+}]_i^{\rm JSR}$, in the vicinity of each specific $i^{\rm th}$
RyR.

The exchange of calcium ions takes place between the RyR clusters, the myoplasm,
and the NSR. In this study, our primary focus is on calcium sparks, which
typically last for approximately 20 ms. We do not consider larger-scale CREs
such as calcium waves, thus limiting the number of calcium sparks and minimizing
the impact on the calcium concentration in the myoplasm $[{\rm Ca}^{2+}]^{\rm
myo}$ and the NSR $[{\rm Ca}^{2+}]^{\rm NSR}$. Consequently, we maintain clamped
concentrations of $[{\rm Ca}^{2+}]^{\rm myo}$  and  $[{\rm Ca}^{2+}]^{\rm NSR}$
at standard values of  0.1 and 1000 $\mu$M, respectively.~\cite{Cheng2008,
Sobie2002} We will also discuss the results obtained with alternative values of
$[{\rm Ca}^{2+}]^{\rm myo}$ and $[{\rm Ca}^{2+}]^{\rm JSR}$.  This approach
aligns with lipid bilayer experiments that utilize clamped calcium
concentrations. Compared to the study of Restrepo et al.~\cite{Restrepo2008}, which incorporated calcium
cycling, our current treatment reduces computational complexity, enabling the
inclusion of the explicit RyR spatial distribution.

The buffering system plays a crucial role in shaping the dynamics of $[{\rm
Ca}^{2+}]_i^{\rm SS}$ fluctuations. In this study, we incorporate the impact of
calcium buffers and specifically examine the effect of CSQ, a buffer protein
found in the JSR, on calcium spark behavior. Taking these factors into
consideration, we can formulate the temporal evolution of $[{\rm
Ca}^{2+}]_i^{\rm SS}$  and $[{\rm Ca}^{2+}]_i^{\rm JSR}$ as follows,
\begin{align}
\frac{d[{\rm Ca}^{2+}]_i^{\rm SS}}{dt} & = J_{i}^{\rm RyR} + J_{i}^{\rm RR} - J_i^{\rm efflux} - J_i^{\rm buffer}, \nonumber\\
\frac{d[{\rm Ca}^{2+}]_i^{\rm JSR}}{dt} & = \frac{V^{\rm SS}}{V^{\rm JSR}} \beta_i \left(-J_{i}^{\rm RyR} + J_{i}^{\rm RR}\right) + J_i^{\rm refill}. 
\label{Eq: master}
\end{align}
The explicit mechanisms are depicted in Fig.~\ref{Fig: schematic}, and detailed
explanations of each flux will be provided in the subsequent subsections.

\subsection{State Transition of RyRs}

RyRs primarily exist in two conformations: open and closed. Numerous studies utilize this two-state model to describe calcium dynamics, where the
transition between these states is predominantly driven by calcium
concentrations.~\cite{Cannell2013, Iaparov2019, Iaparov2021, Walker2014,
Song2016} However, RyRs are also regulated by additional factors, such as
calmodulin and CSQ. To account for these regulatory
complexities, several models have been proposed that expand beyond the two-state
paradigm.~\cite{Colman2022} In our study, we focus primarily on the role of CSQ in
regulating calcium release. Previous work has incorporated the effects of CSQ on RyR transition rates through
${\rm Ca}^{2+}$ binding at cytosolic activation and inactivation sites.~\cite{Shannon2004,
Stern1999} Restrepo
et al.\cite{Restrepo2008} offered a mechanistic explanation for this phenomenon,
later adapted by Sato and Bers into a more comprehensive model with additional
refinements.~\cite{Sato2011, Sato2016} In this model, CSQ binding to RyR
complexes notably reduces RyR activity. To represent this regulation, the model
distinguishes between two RyR states: those bound to CSQ and those unbound. This
approach effectively captures CSQ's modulation of RyR activity and its impact on
calcium release dynamics, as detailed in the work by Restrepo et al.~\cite{Restrepo2008}

\begin{figure}[h!]
  \includegraphics[bb=220 120 620 420,scale=0.4, clip]{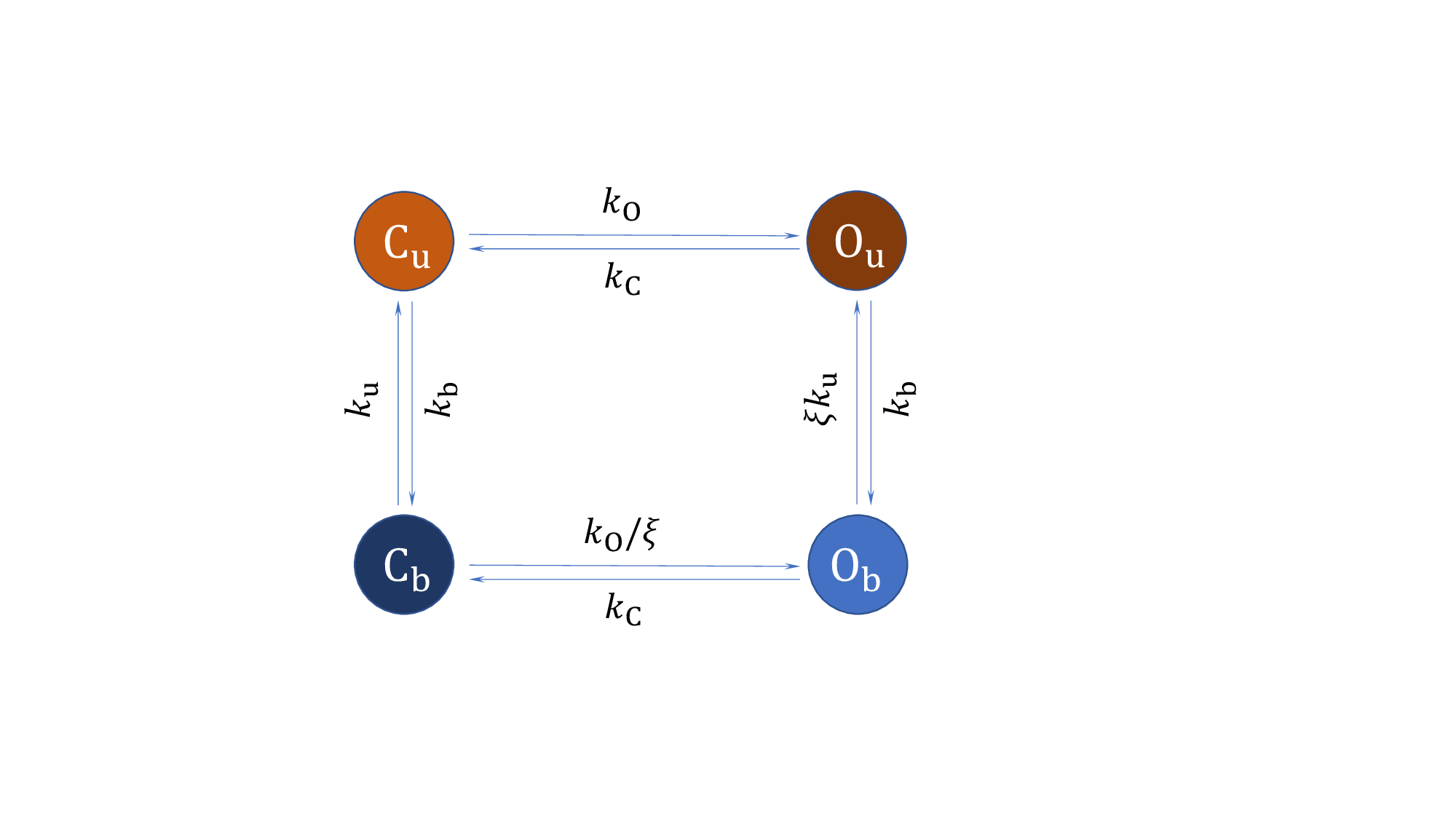}
  \caption{Schematic representation of transitions of four possible states of a RyR channel: closed CSQ-unbound $\rm C_u$,
  open CSQ-unbound $\rm O_u$, closed CSQ-bound $\rm C_b$, and open CSQ-bound $\rm O_b$.}
  \label{Fig: state}
\end{figure}

When $[{\rm Ca}^{2+}]_i^{\rm JSR}$ is elevated, it triggers the release of CSQ
from the RyR complex, which in turn enhances the sensitivity of RyR to $[{\rm
Ca}^{2+}]_i^{\rm SS}$. This phenomenon is shown in Fig.~\ref{Fig: state}, where
two distinct states are illustrated: the CSQ-unbound open state ($\rm O_u$) and the
CSQ-unbound closed state ($\rm C_u$). The opening and closing rates of RyRs, $k_{\rm O}$
and $k_{\rm C}$, are predominantly influenced by the calcium ion concentration in the
subspace $[{\rm Ca}^{2+}]_i^{\rm SS}$, located near the RyR.

\begin{figure*}[htbp!] \includegraphics[bb=0 0 1200 350,scale=0.42,
clip]{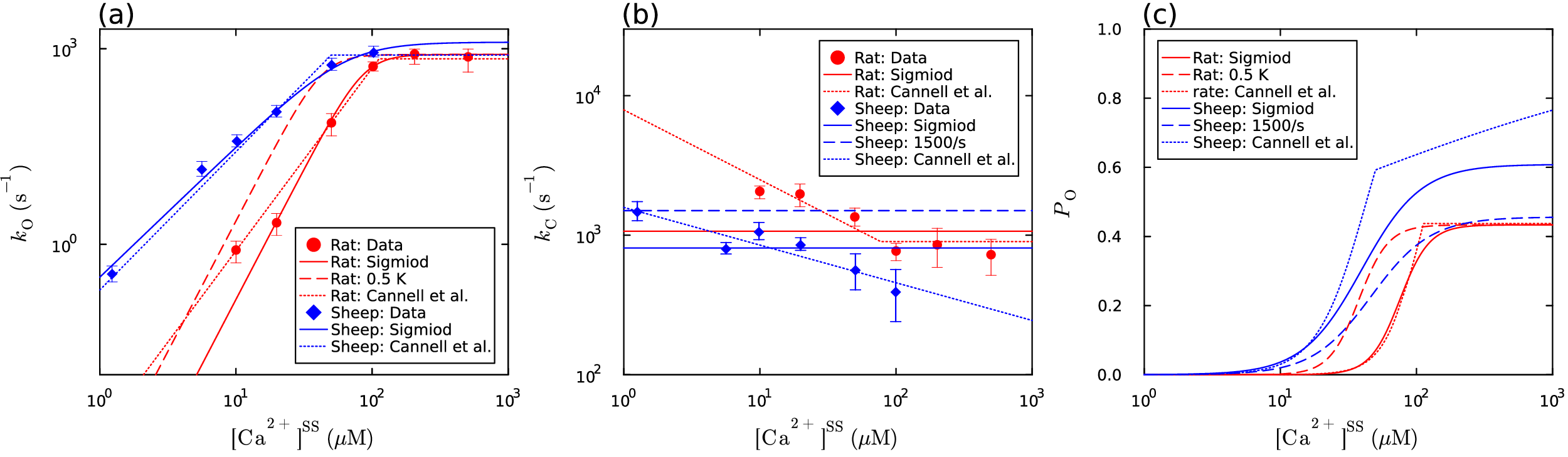} \caption{Transition rates of RyR as a function of calcium
concentration $[{\rm Ca}^{2+}]_i^{\rm SS}$. (a) The opening rate, (b) the
closing rate, and (c) the open probabilities. Data points from Cannell et
al.\cite{Cannell2013} are marked as red circles (rat) and blue diamonds (sheep).
The dotted lines represent the original fits by Cannell et
al.,\cite{Cannell2013} while the solid lines show the new fits obtained in this
study to derive a sigmoid form for RyR open probabilities.  The red dashed line
shows the result of reducing $K$ to half its value for rats, while the blue
dashed line represents the effect of increasing the closing rate to
1500~s$^{-1}$ for sheep.} \label{Fig: rate} \end{figure*}

The opening and closing rates of RyRs, $k_{\rm O}$ and $k_{\rm C}$, have been
determined empirically through single-channel recordings of rat cardiac RyRs
under varying luminal calcium concentration,~\cite{Cannell2013} as reploted in
Fig.~\ref{Fig: rate}(a) and (b). In the original study,~\cite{Cannell2013} the
experimental transition rates were fitted  by the following equations: 
\begin{align} 
k_{\rm O}^{\rm rat} & ={\rm min}\left[a_{\rm O}^{\rm rat}([{\rm Ca}^{2+}]_i^{\rm SS})^{2.8},b_{\rm O}^{\rm rat} \right],\nonumber \\ 
k_{C}^{\rm rat} & ={\rm max}\left[a_{\rm C}^{\rm rat}([{\rm Ca}^{2+}]_i^{\rm SS})^{-0.5},b_{\rm C}^{\rm rat} \right], \label{Eq:rate}
\end{align}
where ``min" and ``max" indicate the selection of the minimum and maximum of the
two values, respectively. The parameters $a_{\rm O}^{\rm
rat}=1.262\times10^{-3}~{\rm s}^{-1}{\rm \mu M}^{-2.8}$, $b_{\rm O}^{\rm
rat}=700~{\rm s}^{-1}$, $a_{\rm C}^{\rm rat}=7906~{\rm s}^{-1}{\rm \mu
M}^{0.5}$, and $b_{\rm C}^{\rm rat}=900~{\rm s}^{-1}$ were obtained from the
experimental data. Similarly, for sheep, the transition rates were fitted as
follows: 
\begin{align} 
  k_{\rm O}^{\rm sheep} & ={\rm min}\left[a_{\rm O}^{\rm sheep}([{\rm Ca}^{2+}]_i^{\rm SS})^{2.12},b_{\rm O}^{\rm sheep} \right],\nonumber \\ 
  k_{\rm C}^{\rm sheep} & =a_{\rm C}^{\rm sheep}([{\rm Ca}^{2+}]_i^{\rm SS})^{-0.27}, 
\end{align} 
where the corresponding parameters are $a_{\rm O}^{\rm sheep}=0.1995~{\rm
s}^{-1}{\rm \mu M}^{-2.12}$, $b_{\rm O}^{\rm sheep}=800~{\rm s}^{-1}$ for
opening rate, and $a_{\rm C}^{\rm sheep}=1582~{\rm s}^{-1}{\rm \mu M}^{0.27}$
for closing rate. These fitted opening and closing rates are presented as dotted
lines in Fig.~\ref{Fig: rate}(a) and (b), respectively.

Typically, the opening rates of RyRs can be expressed using a sigmoid form as: 
\begin{align} k_{\rm O} & =a_{\rm O}\frac{([{\rm Ca}^{2+}]_i^{\rm SS})^n}{([{\rm Ca}^{2+}]_i^{\rm SS})^n+K^n}, 
\end{align} 
It is similar to the formulations used in other studies.~\cite{Sobie2002} However,  
these research~\cite{Sobie2002, Williams2011} introduced a cooperative factor
into the transition rate to describe the interconnection between RyRs. In this
study, we consider the RyRs as nodes in a spatial network, connected through
calcium diffusion as a link (see Section~\ref{Mechanism}B). Therefore, such a cooperative
factor is unnecessary in the current model.  After fitting the data, the
exponential indices $n$ for rat and sheep are found to be 4 and 2, respectively.
The corresponding values of $a_{\rm O}$ are 816.36 and 1259.1, and $K$ values
are 86.96 and 63.35~$\mu$M for rat and sheep, respectively. These sigmoid-fitted
lines are shown as solid lines in Fig.~\ref{Fig: rate}(a), and it can be seen
that both forms of fitting, original ones by Cannell et al. and sigmoid, yield
comparable results.

The closing rates of RyRs, often treated as constants in previous
studies,~\cite{Sobie2002,Walker2014,Williams2011} were similarly fitted as
constants in our analysis, with values of 1066.8s$^{-1}$ for rats and
810.0s$^{-1}$ for sheep. These values differ from those reported by Cannell et
al.\cite{Cannell2013} primarily due to the different fitting methods employed.
This approach not only provides greater flexibility in analyzing the closing
rates but also enables the derivation of a sigmoid form for the open
probabilities. For clarity, we label these constant closing rates as "sigmoid"
in Fig.\ref{Fig: rate}(b).

If we only consider the transitions between two unbound states, the open probability of RyRs $P_{\rm O}$ for a given calcium concentration can be derived from the opening and closing rates as ${k_{\rm O}}/{(k_{\rm O}+k_{\rm C})}$. Specifically, when using the sigmoid-form opening rate and a constant closing rate, the open probability is given by: 
\begin{align} 
  P_{\rm O} & =\frac{a_{\rm O}}{a_{\rm O}+K^n}\frac{([{\rm Ca}^{2+}]_i^{\rm SS})^n}{([{\rm Ca}^{2+}]_i^{\rm SS})^n+\frac{k_{\rm C}K^n}{a_{\rm O}+K^n}}, 
\end{align} 
which clearly exhibit as a sigmoid shape, consistent with
the behavior observed in prior
studies~\cite{Williams2011,Walker2014,Iaparov2021} and shown in Fig.~\ref{Fig:
rate}(c).

For rats, the open probability derived from the sigmoid fit remains close to the
results obtained by Cannell et al.,~\cite{Cannell2013} except for a slight
deviation around 100~$\mu$M. However, for sheep, there is a more significant
discrepancy, primarily due to the assumption of a constant closing rate in our
model. The introduction of a sigmoid-form opening rate allows for more
straightforward exploration of how changes in the rates impact RyR behavior. For
instance, in Fig.~\ref{Fig: rate}, we present a case where the opening rate for
rats is halved as $K=43.48$, which increases the open probability $P_{\rm O}$ at
lower calcium concentrations. Additionally, the impact of a larger closing rate
for sheep is explored, revealing a noticeable reduction in $P_{\rm O}$. These
parameter variations and their effects will be discussed further in the
subsequent sections. 

In vitro experiments have demonstrated that CSQ forms dimers at
calcium concentrations exceeding approximately 500~$\mu$M and higher-order
polymers at concentrations of a few mM.~\cite{Park2004} Typically, the
calcium concentration in the NSR remains high,
maintaining  $[{\rm Ca}^{2+}]_i^{\rm JSR}$ at around 1~mM. RyRs are in
CSQ-unbound states with heightened sensitivity to Ca$^{2+}$ in the
SS. When RyRs are open, calcium ions are released from
the JSR into the SS, resulting in a decrease
in  $[{\rm Ca}^{2+}]_i^{\rm JSR}$.
At low $[{\rm Ca}^{2+}]_i^{\rm JSR}$ levels, CSQ binds to the triadin/junctin
complex, thereby reducing RyR sensitivity.~\cite{Zhang1997,
Gyoke2004, Beard2002} The open CSQ-bound state
($\rm O_b$) and closed CSQ-bound state ($\rm C_b$) are introduced for RyRs with reduced
sensitivity. This state introduces lower
transition rates, denoted as $k_{\rm O}/\xi$, from the closed to the open state, with
$\xi=7.6$.~\cite{Restrepo2008} 

The binding and unbinding rates of CSQ to the RyRs are dependent on the 
$[{\rm Ca}^{2+}]_i^{\rm JSR}$ and the concentration of CSQ, $B_{\rm CSQ}$, given
by,~\cite{Restrepo2008}
\begin{align}
k_{\rm b} &=\hat{M}\tau_{\rm b}^{-1}B_{\rm CSQ}/B^0_{\rm CSQ}, \quad k_{\rm u} =\tau_{\rm u}^{-1},
\end{align}
where the binding and unbinding timescales are $\tau_{\rm b}=5$ ms and $\tau_{\rm u}=125$
ms, respectively. The relative monomer concentration is,
\begin{align}
\hat{M}=\frac{(1+8\rho B_{\rm CSQ})^{1/2}-1}{4\rho B_{\rm CSQ}},
~ {\rm with}~\rho=\frac{\rho_\infty ([{\rm Ca}^{2+}]_i^{\rm JSR})^h}{K^h+([{\rm Ca}^{2+}]_i^{\rm JSR})^h}.
\end{align}
Here, $B^0_{\rm CSQ}=400$ $\mu$M is the normal concentration, ensuring that
$k_{\rm b}$ approaches $\tau_{\rm b}^{-1}$ for low  $[{\rm Ca}^{2+}]_i^{\rm JSR}$. The
parameters in the sigmoid function are chosen as $\rho_\infty=5\times10^3$,
$K=1000$ $\mu$M, and $h=23$.~\cite{Restrepo2008}

The actual release of calcium occurs when the RyR is in an open state $\rm O_u$ and $\rm O_d$,
facilitating the flow of calcium ions from the JSR into the subspace. 
The rate of calcium transfer is determined by the
concentration gradient between these two compartments, expressed as:
\begin{align}
J_{i}^{\rm RyR} & =\frac{i^{\rm RyR}}{2FV^{\rm SS}} \left([{\rm Ca}^{2+}]_i^{\rm JSR}-[{\rm Ca}^{2+}]_i^{\rm SS}\right)\delta_{i,\rm O_u+O_b}.
\end{align}
Here, $J_i^{\rm RyR}$ represents the calcium flux through the RyR and $\delta_{i,\rm O_u+O_b}$ means the RyR $i$
stays in one of two open states $\rm O_u$ or $\rm O_d$. The constant
$i^{\rm RyR}$ is chosen to yield a flux of 0.6~pA for an open RyR with a calcium
gradient of 1~mM.~\cite{Cannell2013, Kettlun2003, Mejia-Alvarez1999} The Faraday
constant, denoted as $F$, is equivalent to 96485 C/mol. The volumes of the
subspace and JSR are assigned values of $V^{\rm SS}=10^{-19}$~L and $V^{\rm
JSR}=2\times10^{-18}$~L, respectively.~\cite{Sobie2002}

\subsection{Spatial network of RyRs}

The CICR mechanism hinges on the diffusion of calcium ions between RyR channels. Upon activation of RyRs, a release of calcium ions transpires, resulting in an increase in calcium concentration and the establishment of a non-uniform distribution of calcium ions within the system. These calcium ions subsequently undergo diffusion driven by concentration gradients as follows,
\begin{align}
  J_{i}^{\rm RR} = \sum_j A_{ij} \left([{\rm Ca}^{2+}]_j - [{\rm Ca}^{2+}]_i\right).
\end{align}
Here, $J_{i}^{\rm RR}$ represents the calcium flux through $i^{\rm th}$ RyR, which is influenced by the difference in calcium concentrations between $j^{\rm th}$ RyR and $i^{\rm th}$ RyR.
The strength of connection between two RyRs is captured by an adjacency matrix,
denoted as $A_{ij}$, within the context of a spatial network. This matrix is
intrinsically tied to the spatial proximity of the RyRs, effectively mirroring
the arrangement of RyR clusters on the JSR. Recent experimental insights,
made possible by super-resolution imaging techniques, have uncovered that RyR
clusters exhibit irregular, non-compact shapes and are randomly distributed
across JSR.~\cite{Baddeley2009,Jayasinghe2018,Shen2019} Therefore, to faithfully
represent the spatial distribution of calcium channels, it is imperative to
generate the RyR network based on experimental data.

The existing literature reveals varied experimental findings regarding cluster
size (the number of RyRs per cluster) and their frequency distribution, as
highlighted in several studies.~\cite{Baddeley2009, Jayasinghe2018, Shen2019}
Generally, smaller clusters are more prevalent than larger ones, following a
power-law distribution.~\cite{Baddeley2009, Jayasinghe2018, Shen2019, Galice2018,
Xie2019} In our previous work,~\cite{Gao2023} we derived a normalized frequency
distribution function based on recent experimental data, expressed as:
\begin{equation} f(x) = 0.991e^{-0.66x} + 0.009e^{-0.017x}, \label{Eq:
cluster_size} \end{equation} where $x$ denotes the cluster size and $f(x)$
represents the frequency distribution of clusters of that size.

When examining a single cluster, the size can be directly assigned. However, for
multiple clusters, the sizes are generated based on the specified frequency
distribution. Considering $N$ RyRs, two random numbers, $r_1 \in [0, 1]$ and
$r_2 \in [0, n_{\rm max}]$, are used. To limit the occurrence of overly large
clusters in a constrained space, the maximum cluster size is capped at $n_{\rm
max} = 100$, consistent with experimental findings that clusters exceeding 100
RyRs are rarely observed.~\cite{Baddeley2009, Jayasinghe2018, Shen2019} If $r_1
< f(r_2)$, $r_2$ is taken as the cluster size; otherwise, both random numbers
are discarded. This procedure repeats until the total number of RyRs equals the
total number of channels, $N$.

In the earlier step, cluster sizes were either randomly determined or explicitly
assigned for a single cluster. The next stage involves generating clusters of
the specified sizes. According to the literature, the structural arrangement of
a cluster is intricate. In this study, we employ the spatial distributions of
RyRs proposed by Jayasinghe et al.,~\cite{Jayasinghe2018} derived from
super-resolution imaging. These findings reveal that clusters are non-compact
and possess irregular geometries. Conveniently, the authors of the referenced
work also outlined a method for the random generation of clusters, which we
directly adopt in this study.

\begin{figure}[h!]
  \includegraphics[bb=0 0 900 510,scale=0.305 ,clip]{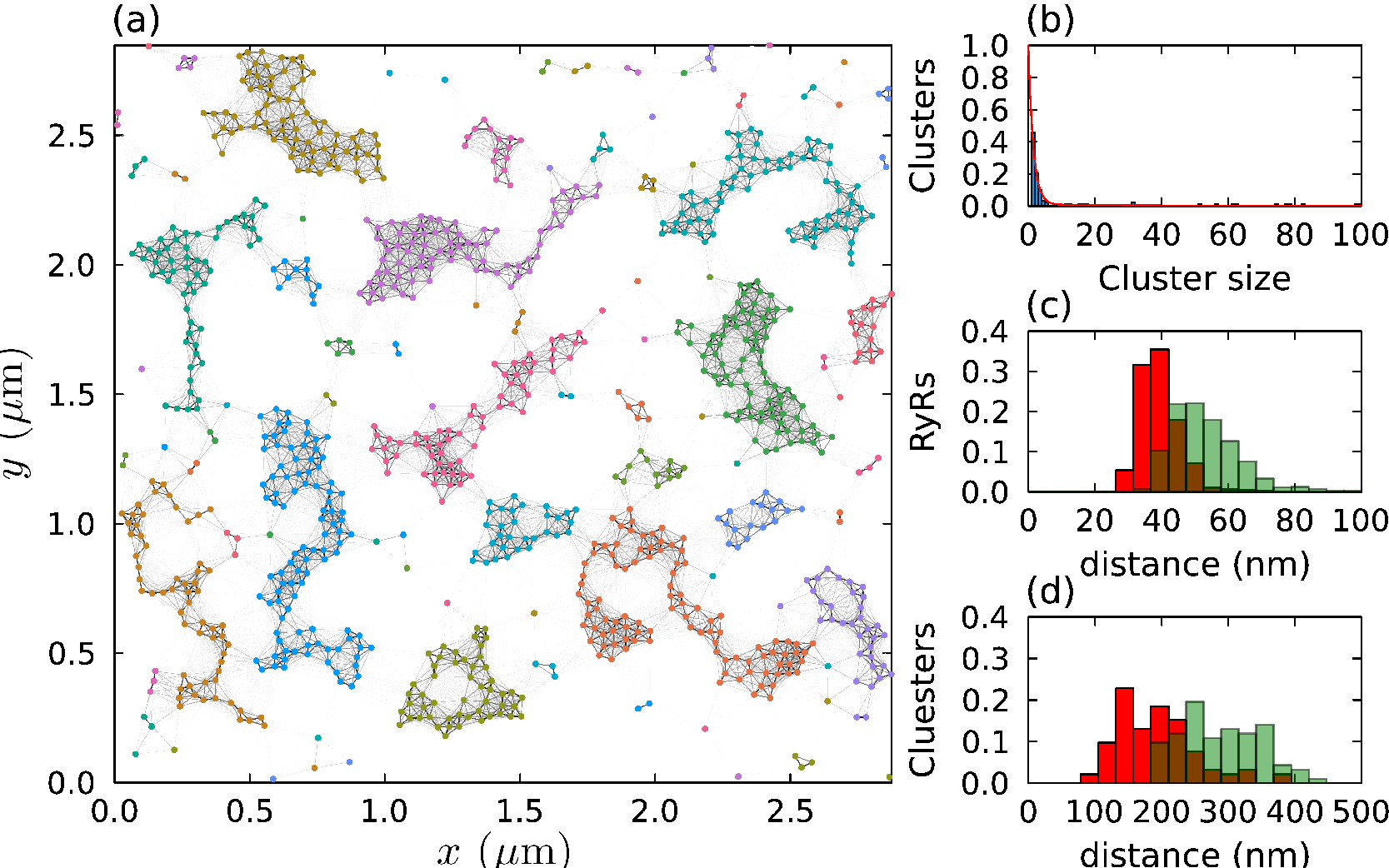}
  
  \caption{ Randomly generated spatial network.  (a) The
  spatial arrangement with distinct colors denoting different clusters of RyRs.
  The width and grayscale of the connecting lines between RyRs indicate the
  strength of their connections as $A_{ij}$.  (b) Comparison of
  randomly generated cluster size distribution (blue histogram) with the
  predetermined distribution (red curve) in Eq.~(\ref{Eq: cluster_size}).  (c)
  Distribution of nearest neighbor distances within clusters: The red bars
  represent the distribution of nearest neighbor distances within clusters
  containing five or more RyRs, while the green line represents the average
  value of the four nearest neighbor distances.  (d) Cluster-level analysis:
  Similar to panel (c) but at the cluster level.  }
  
  \label{Fig: net}
\end{figure}

The simulation starts with an initial position ${\bm x}_1=(0,0)$, which is
stored as the position of the first RyR in a vector. A random direction $\theta$
is then generated, along with a distance $r$ that fluctuates around a mean value
of 40 nm, drawn from a Gaussian distribution with a standard deviation of 7.4
nm. This distribution closely aligns with the observed distance characteristics
in both mean and variance, as reported by Jayasinghe et
al.~\cite{Jayasinghe2018} The position updates to ${\bm x}_2={\bm x}_1+\Delta
{\bm x}$, where $\Delta {\bm x}=(r\cos\theta, r\sin\theta)$ represents the
displacement. The new position ${\bm x}_2$ is added to the vector as the second
RyR's position. This process is iterated by generating additional random
directions and distances, leading to subsequent position updates, such as ${\bm
x}_3$, which are similarly recorded. The procedure continues until the positions
of all RyRs in a cluster of the specified size are stored in the vector. As
illustrated in Fig.\ref{Fig: net}(a), this iterative assembly process results in
the formation of clusters with irregular gaps, resembling those observed in
experiment.~\cite{Jayasinghe2018}

To randomly distribute the clusters with different sizes in a two-dimensional
plane, we follow observations that suggest a random distribution of clusters of
various sizes,~\cite{Baddeley2009, Jayasinghe2018, Shen2019} which is applied
here. A square region of side length $l = d\sqrt{N_{\rm cluster}}$ is
considered, where $d$ represents the average distance between clusters and
$N_{\rm cluster}$ denotes the total number of clusters. Previous studies have
shown that the nearest neighbor distances have a peak near 200 nm with a long
tail.~\cite{Jayasinghe2018} In this work, we adopt a mean distance of 250 nm.

For positioning each cluster, two random values in the range $[0, l]$ are
generated to define its location, and these are added to the positions of the
RyRs in that cluster obtained from the previous step. Since random placement in
the plane may cause overlaps, we address this issue by dividing the square
region into lattices of $30 \times 30$ nm$^2$. As the positions of the RyRs in a
cluster are determined, the corresponding lattices are marked as occupied and
cannot be used again. This method ensures that the clusters are randomly
distributed without overlap, as shown in Fig.~\ref{Fig: net}(a).

To construct the spatial network, we consider RyRs as nodes. The link between
the nodes is abstracted from the calcium diffusion between two RyRs, which
depends on the distance between them. In an unbuffered system, the diffusion of
Ca$^{2+}$ ions is typically proportional to the reciprocal of the distance to
the open RyR, that is $A_{ij} \propto 1/r_{ij}$. However, including calcium
buffers near RyRs reduces the concentration of free Ca$^{2+}$ ions and alters
their distribution.~\cite{Naraghi1997} Therefore, this relationship serves as an
upper bound in formulating Ca$^{2+}$-mediated RyR-RyR
interactions.~\cite{Iaparov2021}
Furthermore, explicit calculations regarding three-dimensional cytoplasmic
calcium propagation indicate that calcium diffusion exhibits an exponential
dependence on distance, as shown in the work by Hernandez-Hernandez et
al.,~\cite{Hernandez-Hernandez2017} and our previous work.~\cite{Jiang2021b} The
rapid decrease with distance $r_{ij}$ in the exponential decay function aligns
with the model of Walker et al., where adjacent RyRs have much stronger
connections than those farther away.~\cite{Walker2014, Walker2015} Hence, we
employ an exponential decay function of the spatial distance $r_{ij}$ between
two RyRs, incorporating two scaling parameters, $\tau_{\rm RR}$ and $r_0$, to
determine the strength and range of connection between RyRs. The mathematical
representation is given by:
 \begin{equation}
  A_{ij}=\frac{e^{-r_{ij}/r_0}}{\tau_{RR}}.\label{Eq: A}
\end{equation}
The range scale $r_0$ is set to 60~nm, roughly equivalent to twice the size of a
RyR. This choice is intended to capture the appropriate interaction range
between RyRs in the simulation, taking into account their physical dimensions.
The parameter $\tau_{\rm RR}$   represents the strength of the connection
between two RyRs, which in turn reflects the magnitude of the flux or exchange
of ions between them. However, this parameter cannot be directly estimated from
available experimental data. Given the calcium diffusion constant $D_{\rm Ca} =
200~\mu{\rm m}^2/{\rm s}$, we estimate the diffusion time as $\tau_{RR} =
{r_0^2}/{2D_{\rm Ca}} \approx 0.01~{\rm ms}$.

Although our primary focus is on a single cluster of RyRs, we now aim to present
a spatial network illustrating the randomized distribution of RyRs across
multiple clusters. This network provides a clearer representation of clusters
with varying sizes. As an illustrative example, Fig.~\ref{Fig: net} shows a
network of 1000 RyRs, successfully replicating the cluster structure observed in
previous studies. Different colors are used in the figure to distinguish between
the RyR clusters, which exhibit random shapes and sizes. Notably, by adopting
the self-assembly process,~\cite{Jayasinghe2018} the clusters show
larger gaps, similar to those observed and reported in that reference.

To better understand the spatial distribution of RyRs and evaluate the
consistency between our simulations and experimental data, we analyze the
distribution of cluster sizes, as shown in Fig.~\ref{Fig: net}(b). This analysis
reveals that our model closely adheres to the predetermined cluster size
distribution given by Eq.~(\ref{Eq: cluster_size}), which is based on
experimental data.~\cite{Shen2019} Additionally, we examine the spatial
relationships between RyRs within clusters by analyzing the nearest neighbor
distances. Fig.~\ref{Fig: net}(c) illustrates the distribution of these
distances, along with the average of the four nearest neighbor distances. The
strong agreement of these metrics with experimental data underscores the
accuracy of our model in capturing the spatial arrangement of RyRs.
Furthermore, we extend this analysis to explore the distribution of distances
between clusters, which follows a long-tail distribution consistent with
experimental observations,~\cite{Jayasinghe2018} as depicted in Fig.~\ref{Fig:
net}(d).

In our model, each RyR is represented as a colored dot in Fig.~\ref{Fig: net}(a)
and is abstracted as a node within the spatial network. The connections between
RyRs are visually depicted as lines, with the width and grayscale of these lines
corresponding to the strength of the connections as $A_{ij}$ in Eq.~(\ref{Eq: A}). Notably,
the stronger connections within clusters are clearly visible, highlighting their
critical role in facilitating calcium release and generating calcium sparks.

\subsection{Exchange of calcium and buffering}

In our study, we clamp the calcium concentrations in NSR and myoplasm at
constant levels. The resting calcium concentration in the myoplasm, denoted as
$[{\rm Ca}^{2+}]^{\rm myo}$, is approximately 0.1~$\mu$M, as determined
experimentally.~\cite{Cheng2008, Sobie2002} The efflux of calcium ions from the
subspace to the myoplasm, represented by $J_i^{\rm efflux}$, occurs due to the
concentration gradient between the SS and myoplasm as,~\cite{Sobie2002,Williams2011,Restrepo2008}
\begin{align} J_i^{\rm efflux} = \frac{[{\rm Ca}^{2+}]_i^{\rm SS} - [{\rm
Ca}^{2+}]^{\rm myo}}{\tau^{\rm efflux}},  \end{align} where, $\tau_{\rm
efflux}$ is set as $0.02$~ms, which corresponds to diffusion over a distance
of approximately 80~nm, the average distance required to exit the proximal
space, which has a cylindrical shape with a radius of 200~nm.~\cite{Restrepo2008}

As calcium is released from the JSR through
the RyRs, the $[{\rm Ca}^{2+}]_i^{\rm JSR}$ decreases, initiating calcium refilling from the NSR, where
the calcium concentration is maintained at 1~mM.~\cite{Cheng2008, Sobie2002,
Williams2011, Walker2015} The refill flux ($ J_i^{\rm refill}$) is proportional
to the concentration gradient between the NSR and the JSR,~\cite{Sobie2002,Williams2011,Restrepo2008}
\begin{align}
J_i^{\rm refill} & =\frac{[{\rm Ca}^{2+}]^{\rm NSR}-[{\rm Ca}^{2+}]_i^{\rm JSR}}{\tau^{\rm refill}},
\end{align}
where the time constant $\tau_{\rm refill}$ is chosen as 5~ms.~\cite{Restrepo2008}

Within the framework of calcium regulation, calcium ions also interact with
calcium buffers, including sarcolemmal (SL) and sarcoplasmic reticulum (SR)
membrane buffers, as well as the versatile protein CaM. The
characteristics of these buffers can be mathematically described by the equation,
\cite{Smith1996, Smith1998}
\begin{align}
J_i^{\rm buffer} = -k_{\rm on}[{\rm B}][{\rm Ca}^{2+}]_i + k_{\rm off}([{\rm B}]_T - [{\rm B}]_i).
\end{align}
In this equation, the rate at which calcium ions bind to the buffer is denoted
by $k_{\rm on}$, while the rate at which they dissociate from the buffer is
represented by $k_{\rm off}$. Moreover, $[{\rm B}]_T$ and $[{\rm B}]_i$ signify
the total concentration of buffer and the concentration of unbound buffer,
respectively. Please refer to Table~\ref{Buffering parameters} for specific
values of the constants related to buffering.

\renewcommand\tabcolsep{0.42cm}
\renewcommand{\arraystretch}{1.2   }
\begin{table}[h!]
  \caption{Buffering parameters.~\cite{Smith1998}
    \label{Buffering parameters}}
  \begin{tabular}{crrr}\bottomrule[1.5pt]
    Buffer & $[{\rm B}]_T$ ($\mu$M) & $k_{\rm on}$ ($\mu$M$^{-1}$s$^{-1}$) & $k_{\rm off}$ (s$^{-1}$) \\\hline
    CaM    & 24    & 100                 & 38      \\
    SR     & 47    & 115                 & 100     \\
    SL     & 1124  & 115                 & 1000    \\
   \bottomrule[1.5pt]
  \end{tabular}
\end{table}

In this research, we concentrate primarily on the modulation of calcium release
mediated by CSQ, the principal calcium buffer residing within the JSR. To model
this process, we adopt the fast buffering approximation.~\cite{Wagner1994} While
this approximation does not strictly conserve mass, it is acceptable in our
context because the calcium concentration is clamped, making mass conservation
irrelevant. The $\beta_i$ factor, appearing in Eq.~(\ref{Eq: master}), is
defined as per the formulation presented in Restrepo's study,
\cite{Restrepo2008}
\begin{align}
\beta_i(c_i)=\left(1+\frac{K_CB_{\rm CSQ}n(c_i)
+\partial_{c_i}n(c_i)(c_iK_c+c_i^2)}{(K_C+c_i)^2}\right)^{-1},\label{Eq: beta}
\end{align}
were $K_{C}=600~\mu$M, and $c_i$ represents the luminal $[{\rm Ca}^{2+}]_i^{\rm
JSR}$. The total buffering capacity $n(c_i)$ is defined as
$n(c_i)=\hat{M}(c_i)n_M+(1-\hat{M}(c_i))n_D$, where $n_M=15$ and $n_D=30$ represent
the capacity for monomeric and dimeric CSQ, respectively.~\cite{Park2004} 

\section{Simulations of calcium release event}\label{results}

\subsection{Trigger and stochastic algorithm}
In this section, we perform simulations of CREs based
on the calcium regulation mechanism described in the preceding section. We
examine both spontaneous CREs and LCC-evoked CREs, each initiated by distinct
mechanisms.

For spontaneous CREs, RyRs open randomly due to a very small opening
probability of RyRs under resting calcium concentrations in physiological
conditions or their malfunction in pathological scenarios. The open and close
rates are governed by Eq.~(\ref{Eq:rate}). Under resting conditions, characterized by
an approximate $[{\rm Ca}^{2+}]_i^{\rm SS}$ of 0.1~$\mu$M, 
RyRs predominantly remain closed. An increased ratio between open and
closing rates enhances spontaneous RyR openings, leading to the experimental
observation of spontaneous sparks.~\cite{Cannell2013} However, these spontaneous
sparks are infrequent, making it impractical to comprehensively study CRE
behavior using this approach. In our simulations, we employ the rates described
in Eq.~(\ref{Eq:rate}) to mirror realistic conditions, despite yielding negligible
sparks at low $[\rm Ca^{2+}]_i^{\rm SS}$. Instead, we initiate spontaneous CREs
by briefly placing a RyR in the open state for a duration of 0.001 ms.

The activation of LCCs is a critical trigger for CREs,
responding to various stimuli such as changes in voltage, hormonal signals, and
signaling molecules. While the dynamics of LCC opening are complex, our focus
here is on how LCCs evoke calcium sparks and influence their formation and
characteristics. For simplicity, we consider only  events with single LCC release; in
this case, the formation of a calcium spark is not influenced by
multiple LCC releases. Further details will be presented in
Section~\ref{results}C.

In our study, we utilize a spatial network model where RyRs are
considered as nodes within the network. To simulate LCC-evoked CREs, we
strategically position LCC calcium release sites in close proximity to RyR
nodes, eliminating the need to introduce additional nodes to represent LCCs. The
calcium flux is maintained at a constant rate of 0.5~pA, and the duration of LCC
opening is consistently set at 6.7~ms, a value derived from the time constant
associated with sparklet-spark coupling latency.~\cite{Wang2001} In mode LCC I,
only a single LCC release is assigned to a cluster. In mode LCC II, as suggested 
by experimental observations, LCCs are roughly
one-seventh in number compared to the RyRs within a cluster. This configuration
ensures the activation of calcium sparks.~\cite{Bers1993} Other release type will
be also discussed in Section~\ref{results}C.

In this study, we employ a stochastic simulation methodology with fixed time
steps. At each predetermined interval, for every transition between two of the
four states of every RyR, a random number is generated to determine whether the
transition occurs, based on the corresponding transition rates. Simultaneously,
calcium concentration is updated in line with the flux dynamics described in the
previous section. To ensure accuracy, we use a time step of 0.001~ms, which
minimizes the likelihood of multiple reactions occurring within the same time
step.

\subsection{Exemplary CREs within a single cluster}

\begin{figure*}[htbp!]
  \includegraphics[bb=0 7 500 400, scale=0.5,clip]{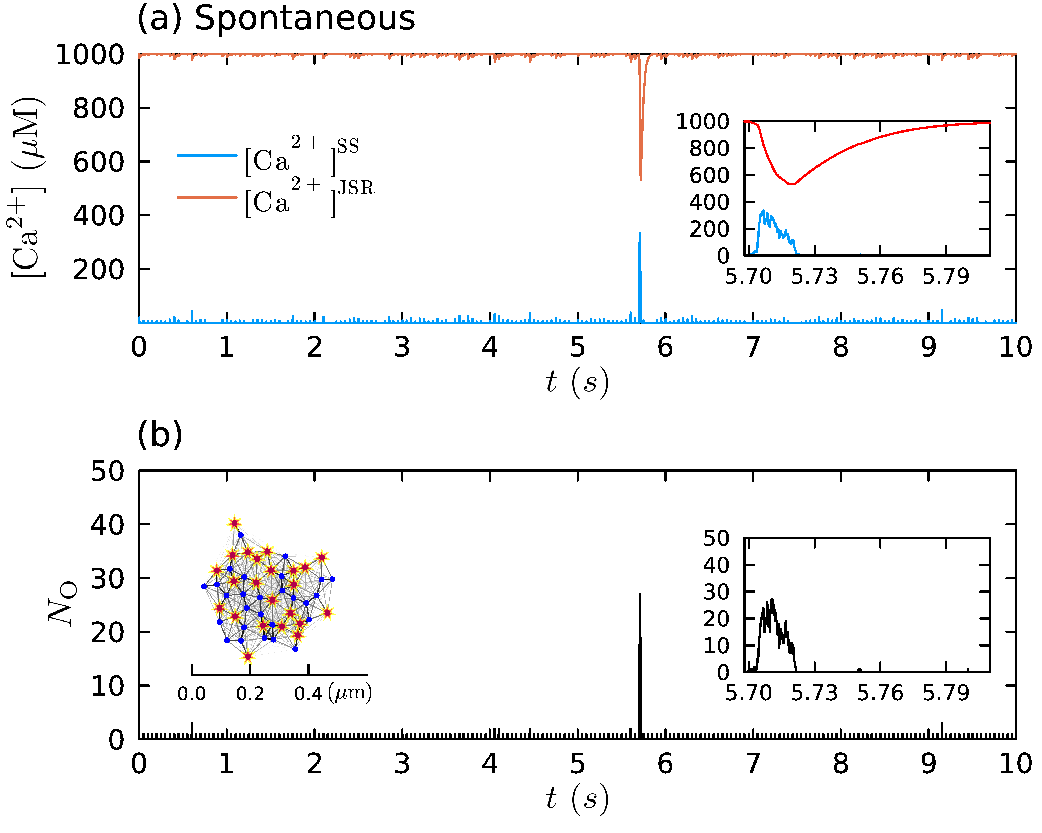}
  \includegraphics[bb=0 7 500 400, scale=0.5,clip]{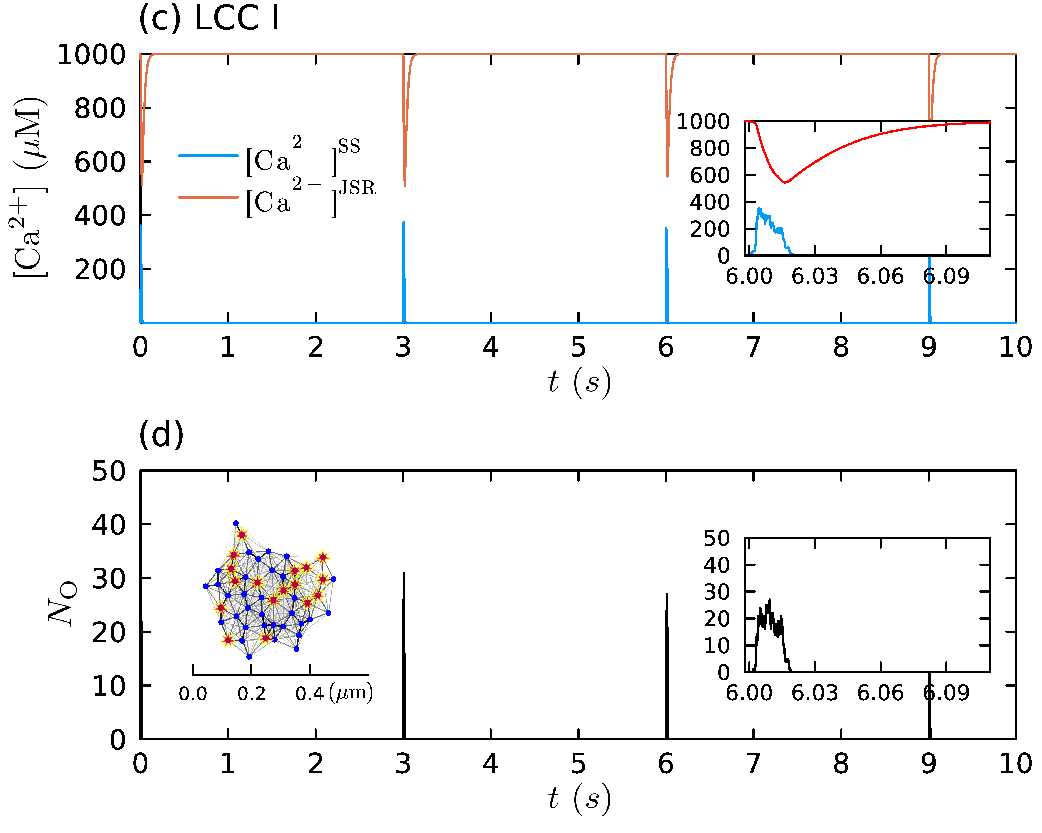}
  \caption{Temporal calcium release profile in a single cluster of 50 RyRs for
  the spontaneous scenario (panels a, b) and LCC I scenario (panels c, d). (a,
  c) Depicts the temporal evolution of mean calcium concentrations, including $[{\rm
  Ca}^{2+}]^{\rm SS}$ (blue curves) and $[{\rm Ca}^{2+}]^{\rm JSR}$ (red
  curves). (b, d) show the variation in the number of open RyR channels $N_O$ over
  time. The right subfigure highlights the CREs near
  $t=5.7$ and 6.0~s for the spontaneous (a, b) and LCC I (c, d) scenarios,
  respectively. The subfigure on the left illustrates the spatial network of the
  analyzed cluster with opened RyRs (red stars with yellow borders) at 5.71 and
  6.01s for spontaneous and LCC I scenarios, respectively. } \label{Fig:
  Single0} 
\end{figure*}

In this section, we present examples of spontaneous and LCC-evoked CREs within a
cluster consisting of 50 RyRs over a simulation period of 10~s, with triggers
occurring every 0.05~s for spontaneous scenarios and every 3~s for LCC-evoked
scenarios. The temporal evolution of mean calcium concentrations, $[\rm
Ca^{2+}]^{\rm SS}$ and $[\rm Ca^{2+}]^{\rm JSR}$, and the number of open RyRs
$N_{\rm O}$ in the cluster is illustrated in Fig.~\ref{Fig: Single0}. Additionally, the
figure includes a depiction of the spatial structure of the analyzed cluster,
generated using the prescribed method.

In the spontaneous scenario, the simulation reveals that most rapidly opening
RyRs tend to close soon after and are unable to trigger further opening of other
RyRs. Only a few open RyRs can initiate a CRE involving more than one RyR,
resulting in short-lived small peaks, characteristic of calcium quarks. Notably,
we observe a distinct CRE that induces significant fluctuations in the mean
calcium concentration $[{\rm Ca}^{2+}]^{\rm SS}$ around 5.7~s with open RyRs
much more than the calcium quarks. This concentration increases from its resting
value of approximately 0.1~$\mu$M to around 300~$\mu$M. Concurrently, the $[{\rm
Ca}^{2+}]^{\rm JSR}$ decreases from 1 mM to approximately 500~$\mu$M, which is
consistent with findings in the literature.~\cite{Cannell2013, Walker2014,
Walker2015, Sobie2002, Williams2011} Obviously, the ${\rm Ca}^{2+}$ in JSR is
not depleted.  LCCs prove highly effective at triggering CREs. In an
illustrative simulation [see Fig.~\ref{Fig: Single0}(c) and (d)], all four LCC
triggers evoke CREs with amplitudes similar to those in the spontaneous
scenario. These CREs, with long duration and many RyRs involved, align well with
experimentally observed spontaneous and LCC-evoked calcium
sparks.~\cite{Cheng1993, Wang2001}

The accompanying subfigures offer a detailed representation of the CRE occurring
near time $t$=5.7 and 6.0~s for the spontaneous and LCC-evoked scenarios, respectively.
Fig.~\ref{Fig: Single0}(b) illustrates that when a specific RyR is set to the open
state, it remains open for roughly 5~ms. During this interval, only this RyR
remains open, while the other RyRs stay closed as the 
[$\rm Ca^{2+}]^{\rm SS}$ increases slowly. After approximately 5~ms, other RyRs
are rapidly activated, and the number of open RyRs increases quickly to around
25, representing about half of the total RyRs in the cluster. A similar rapid
increase in the number of open RyRs is observed at about 5~ms following LCC
release in Fig.~\ref{Fig: Single0}(d). At the outset of the release, no RyRs are
opened, although the  [$\rm Ca^{2+}]^{\rm SS}$ increases
gradually, akin to the spontaneous case. The LCC release serves a role similar
to that of a long-lived open RyR in evoking CREs. The number of open RyRs
remains at about 20 and then decreases rapidly within several milliseconds,
while the  [$\rm Ca^{2+}]^{\rm SS}$ continuously decreases
to its resting value. Both of these processes last approximately 20~ms. The time
evolution of RyRs and the calcium concentrations in both spontaneous and
LCC-evoked cases exhibit similar behaviors, arising from the shared mechanism
post-evocation.

\subsection{CREs on clusters with different sizes}

In our preceding discussion, we provided an illustrative example of calcium
release within a cluster consisting of 50 RyRs. Triggered by the spontaneous
opening of a RyR or LCC calcium release, the cluster functions as an on-off
switch, producing calcium quarks with very short lifespans and only a few
activated RyRs, or calcium sparks with much longer lifespans and approximately
half of the RyRs opened. To gain a broader understanding, we conducted
simulations with multiple randomly generated clusters of various sizes,
including $N_{\rm RyR}=30$, 50, and 80.

For the spontaneous scenario, we generated 5000 spatial distributions and their
corresponding networks and triggered them 10 times each. As seen in the above
examples, LCCs are more effective at evoking calcium sparks. Therefore, we only
generated 500 spatial distributions for the LCC-evoked scenario. Additionally,
we considered multicluster cases with 1000 RyRs, with 200 and 20 spatial
distributions considered for spontaneous and LCC-evoked scenarios, respectively.

\begin{figure}[htbp!] \includegraphics[bb=7 7 785
950,scale=0.31,clip]{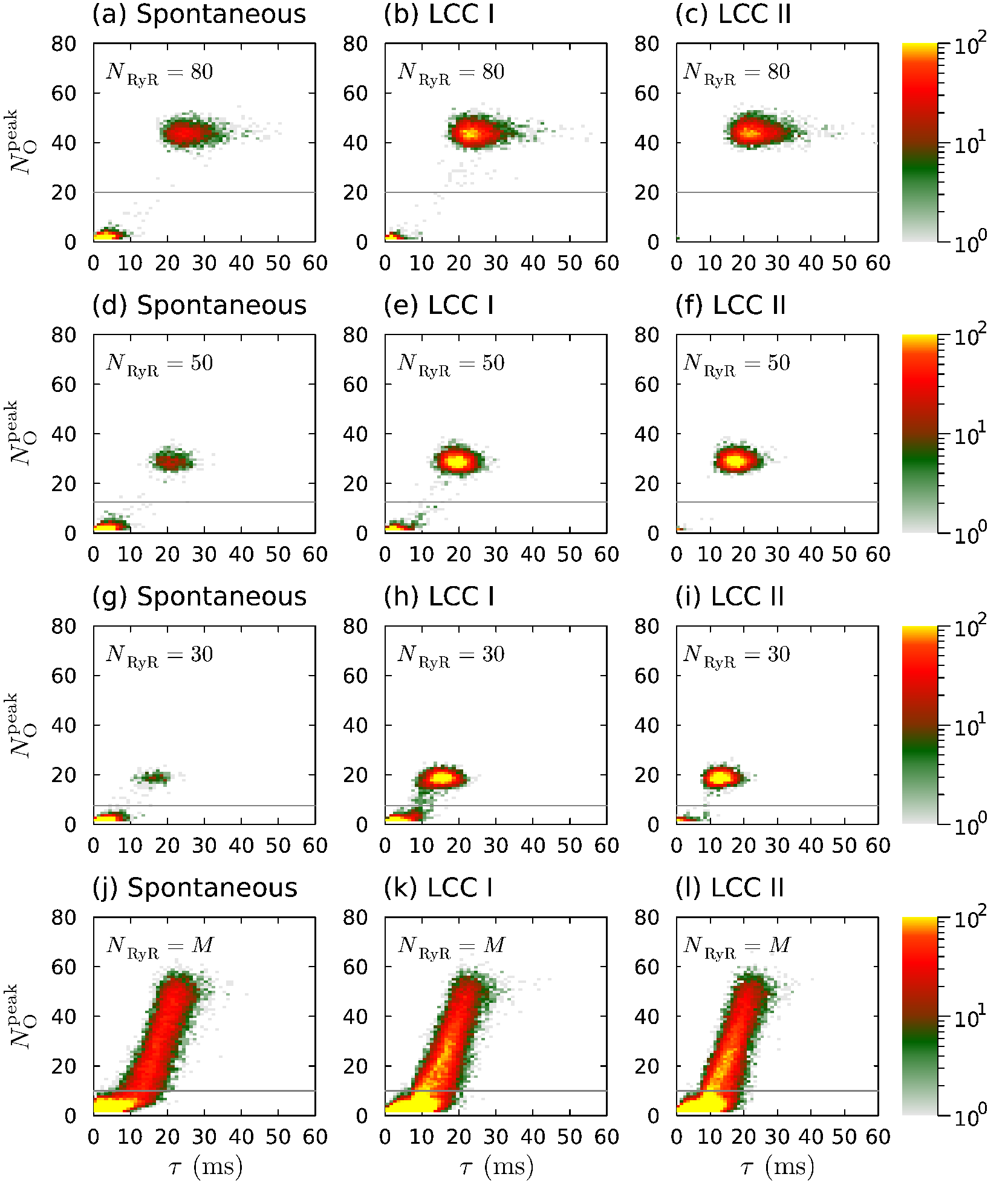}
\caption{2D histograms representing the distribution of CREs
on clusters with sizes 80, 30, and 50, as well as multicluster cases (M), against
the number of peak open RyRs $N_{\rm O}^{\rm peak}$ and duration $\tau$, with
bin sizes of 1 and 1 ms, respectively. Horizontal lines in panels (a-i) and
panels (j-l) indicate $N_{\rm O}^{\rm peak}$ values of $N_{\rm RyR}/4$ and 10,
respectively.} \label{Fig: Single} \end{figure}

In Fig.~\ref{Fig: Single}, we present the distribution of the CREs against the
peak number of open RyRs, denoted as $N_{\rm O}^{\rm peak}$, and the duration of
these CREs, represented as $\tau$, for different cluster sizes. As shown in
Fig.~\ref{Fig: Single0}, the activation of a cluster occurs rapidly in the
spontaneous scenario, primarily driven by a long-lived RyR channel. To
facilitate analysis, we defined the total duration, $\tau$, as the time from the
first RyR opening to when all RyRs within the cluster have closed. In the
LCC-evoked scenarios, the calcium concentration begins to rise upon LCC opening,
marking the onset of the calcium spark. In this case, we defined the initiation
time point at the LCC opening to facilitate a direct comparison with spontaneous
CREs.

Remarkably, a clear demarcation emerges when distinguishing CREs
characterized by peak open RyRs $N_{\rm O}^{\rm peak}$ and
duration $\tau$ in both scenarios, as shown in Fig.~\ref{Fig: Single}(a-i). To
elucidate this, consider the case of an 80 RyR cluster where we can establish a
critical threshold at $N_{\rm O}^{\rm peak}$ $=$ $20$ in Fig.~\ref{Fig:
Single}(a-c). This threshold effectively segregates CREs into two distinct
categories. Those above the threshold tend to converge around $N_{\rm O}^{\rm
peak}$ $=$ $40$ with a corresponding $\tau$ of approximately 25~ms. These larger
CREs, characterized by larger $N_{\rm O}^{\rm peak}$ and $\tau$, are indicative
of calcium sparks. They release a more substantial amount of calcium and are
more likely to align with experimental observations. Conversely, CREs with
smaller $N_{\rm O}^{\rm peak}$ possess a duration of less than 10~ms with few
RyRs, corresponding to calcium quarks. Such results confirm that clusters of
RyRs function as on-off switch.

The 2D histogram shown in Fig.~\ref{Fig: Single}(a, d, g) for the spontaneous CREs
exhibits a similar shape to the histograms in Fig.~\ref{Fig: Single} (b, e, h) for
LCC-evoked CREs with initiation mode I. It is consistent with the observed
similarity of the two types of calcium sparks in experiments.~\cite{Wang2001}
Moreover, LCC release is much more effective at evoking CREs than rapidly
opening RyRs. As depicted in Fig.~\ref{Fig: Single}, the number of CREs in the
LCC-evoked scenario I, even with fewer triggers, is significantly larger than in
the spontaneous case. We also adopted mode II in the LCC-evoked scenario, which
aligns with experimental findings suggesting that LCCs number approximately 1/7
of the RyRs within a cluster, ensuring the activation of calcium sparks.~\cite{Bers1993}

To provide a clearer understanding of the properties of CREs with respect to cluster size, we also present reference results with
multicluster configurations in Fig.~\ref{Fig: Single}(j-l).  In the current work,
the cluster size obeys a power-law distribution following the experimental
results.~\cite{Baddeley2009,Jayasinghe2018,Shen2019,Galice2018,Xie2019}
The frequency of clusters with a size larger than 10 is smaller and decreases
slower than those with a size smaller than 10. The results suggest that the time
scale of CREs for different cluster size is about 20~ms, while $N_{\rm O}^{\rm
peak}$ has a wide spread from 0 to about 50. When combined with the results from
a single cluster, it is evident that the amplitudes of calcium sparks are
determined by the cluster size, while the duration is less sensitive to the
cluster size.

\subsection{Calcium sparks with different LCC triggers}

In this work, we consider  triggers with a single release from LCCs as a
constant calcium flux to evoke calcium sparks. However, the opening dynamics of
LCCs are much more complex. For instance, LCCs typically have multiple states
and transition between them, and several detailed mechanisms have been developed
to describe LCC opening and the corresponding calcium flux.~\cite{Hinch2004,
Jafri1998, Shannon2004, Mahajan2008, Wei2021, Restrepo2008} Our focus here is
on how LCCs evoke calcium sparks and the formation and characteristics of these
sparks. Therefore, we only consider events with a single LCC release. In this
case, the formation of a calcium spark is not influenced by multiple LCC
releases. This approach is reasonable for in vitro experiments studying the
formation and properties of individual calcium sparks and is also applicable to
some physiological conditions, as well as certain pathological scenarios.

\begin{figure}[h!] \includegraphics[bb=0 7 790
  460,scale=0.31,clip]{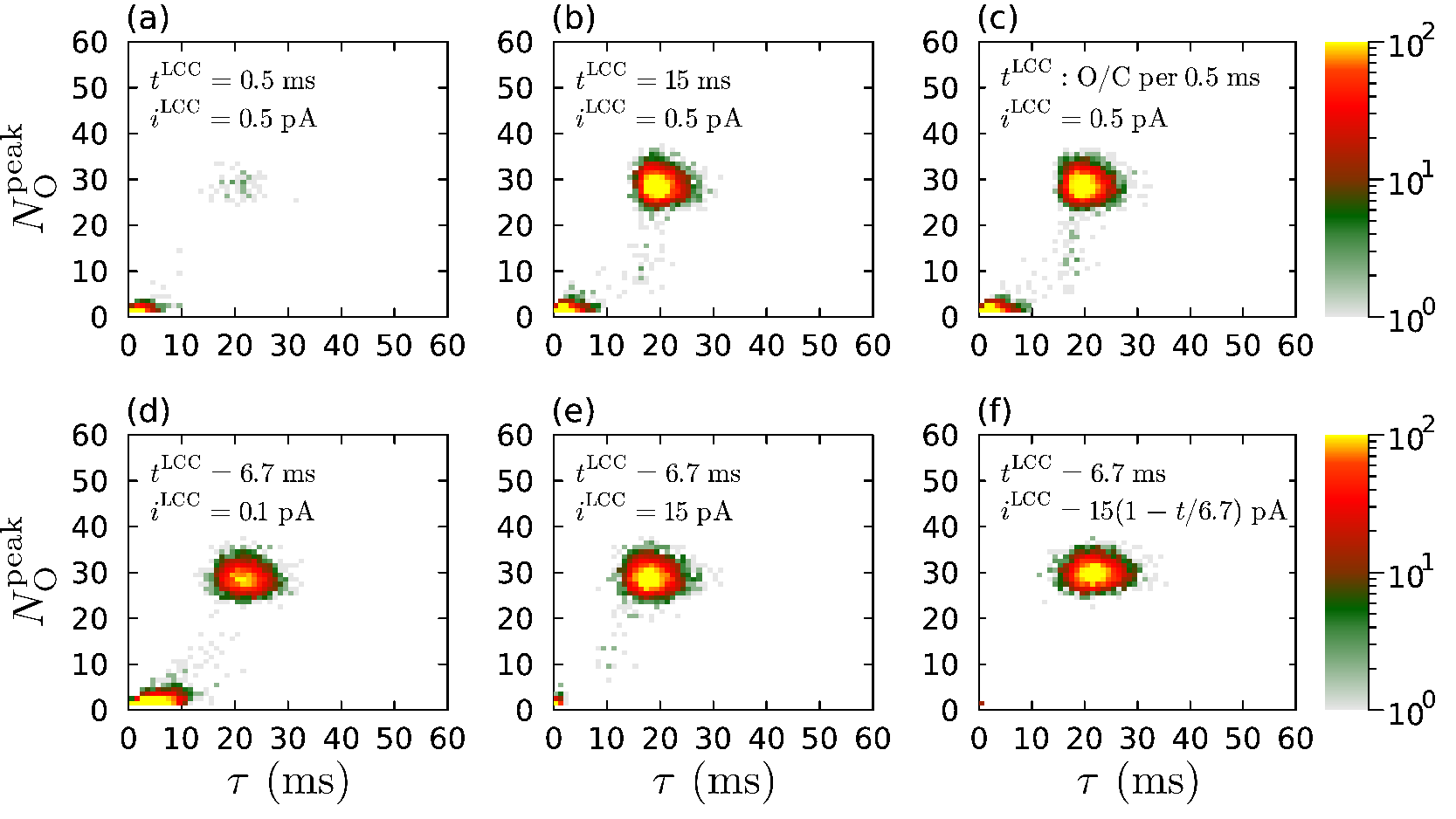}
  
  \caption{2D histograms representing the distribution of CREs on clusters of
  size 50 with different LCC triggers. Pannels (a), (b), and (c) show a constant flux
  $i^{\rm LCC}=0.5$ pA with durations $t^{\rm LCC}$ of 0.5 ms, 15 ms,
  and an interval of opening and closing every 0.5 ms over a total period of 15
  ms, respectively. Pannels (d), (e), and (f) show a fixed LCC duration of 6.7 ms, with varying
  flux strengths: $i^{\rm LCC} = 0.1~\text{pA}$ , $i^{\rm LCC} =
  15~\text{pA}$, and $i^{\rm LCC} = 15(1 - t/6.7)~\text{pA}$, respectively.}\label{Fig: LCCti} 
  
\end{figure}

Typically, during LCC opening and calcium release, the calcium flux is treated
as constant.~\cite{Sobie2002,Williams2011} However, the duration and magnitude
of this flux can vary. For example, as noted in previous studies, the opening
duration of LCCs can range from approximately 0.1~ms to around 20~ms.~\cite{Wang2001} 
Here, we will discuss whether such differences in opening
time and flux constant of LCC affect the formation and characteristics of
calcium sparks. The results of this analysis with a cluster of 50 RyRs as
example are presented in Fig.~\ref{Fig: LCCti}.

In the  calculations in above sections, we considered an LCC release with a flux constant of
$i^{\rm LCC} = 0.5~\text{pA}$ and a duration of 6.7~ms. Here, we keep the flux
constant at 0.5~pA and varied the duration to 0.5~ms, which is typical of LCC
opening times under physiological conditions, and to 15~ms, a longer duration
observed under experimental conditions,~\cite{Wang2001} as shown in
Fig.\ref{Fig: LCCti} (a) and (b). Additionally, we examined a case where the LCC
opened and closed at intervals of 0.5~ms over a total period of 15~ms, as
presented in Fig.~\ref{Fig: LCCti} (c). The results show that longer LCC
durations increase the occurrence of calcium sparks, while the peak number of
open RyRs $N_{\rm O}^{\rm peak}$ and the duration $\tau$ of the calcium sparks remain largely unaffected by
the LCC duration.

In Fig.~\ref{Fig: LCCti} (d-f), we keep the duration time of LCC fixed at 6.7~ms and
vary the flux cosntant, considering cases with $i^{\rm LCC} = 0.1~\text{pA}$
and $i^{\rm LCC} = 15~\text{pA}$, as well as a linearly decreasing flux modeled
as $i^{\rm LCC} = 15(1 - t/6.7)~\text{pA}$. The results indicate that increasing
the flux strength significantly raises the likelihood of calcium spark
initiation. However, despite this increase in spark probability, the general
properties of the calcium sparks, such as their duration and amplitude, remain
consistent across different flux strengths.

These results suggest that if a calcium spark is triggered by a single release
from LCCs, its properties remain largely independent of the specific trigger
type. The differences in the type of trigger mainly influence the likelihood of
calcium spark occurrence.  Thus, the explicit mechanisms governing LCC state
transitions impact the general scale of RyR release and, consequently, the
overall calcium release within the cell. However, in this study, we focus solely
on the probabilities associated with individual calcium sparks. Therefore, in
the following calculations, we consider only an LCC release with a constant flux
of $i^{\rm LCC} = 0.5~\text{pA}$ and a duration of 6.7~ms.

\section{Regulation of clacium sparks }\label{Regulation}

In previous sections, we explored the formation of calcium sparks and observed
that their peak number of open RyRs and duration remain consistent, whether
evoked spontaneously or triggered by various types of LCC release. In this
section, we examine the regulation of calcium sparks by analyzing the influence
of RyR transition rates, calcium concentrations in the NSR and myoplasm, and the
role of CSQ. 

\subsection{Regulation  by RyR transition rates}

Cannell et al. highlighted significant differences in RyR behavior between rats
and sheep.~\cite{Cannell2013} As illustrated in Fig.~\ref{Fig: rate}, the
transition rates of RyRs differ between species. While it is clear that the
magnitude of the RyR transition rate influences the likelihood of calcium spark
occurrence, the question remains: does it also impact the characteristics of the
calcium sparks themselves? To explore this, we present in Fig.~\ref{Fig:
plotrate} the results using different transition rates with 10000 simulations
under a LCC trigger, investigating their potential effects on calcium spark
properties.

\begin{figure}[htbp!] \includegraphics[bb=0 7 795 470,scale=0.31,clip]{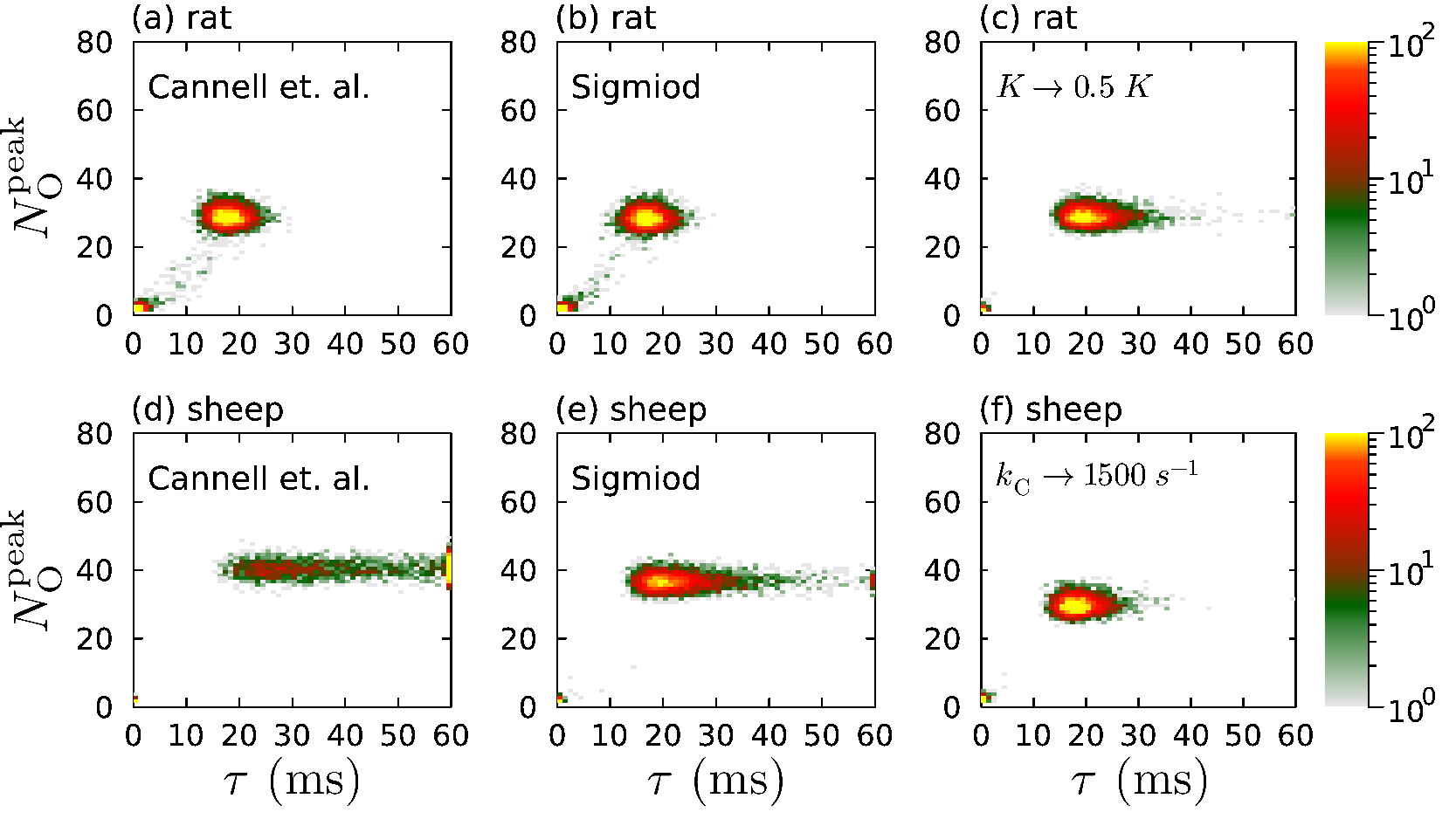}
  \caption{2D histograms representing the distribution of CREs on clusters of size 50 with varying RyR transition rates. (a) Transition rates for rat RyRs as fitted by Cannell et al. (b) Transition rates for rat RyRs fitted using the sigmoid form. (c) Same as (b) but with $K$ reduced to 0.5$K$. (d) Transition rates for sheep RyRs as fitted by Cannell et al. (e) Transition rates for sheep RyRs fitted using the sigmoid form. (f) Same as (e) but with the closing rate increased to 1500~s$^{-1}$.}\label{Fig: plotrate} 
\end{figure}

In Fig.~\ref{Fig: plotrate} (a), we present the results using the transition
rates for rat RyRs as fitted by Cannell et al., shown as the dotted line in
Fig.~\ref{Fig: rate}. These rates have been applied in the previous
calculations. The results show a clear concentration of CREs. When the
transition rates are adjusted to those fitted using the sigmoid form for the
opening rate and a constant closing rate, the calcium spark properties, including the
peak number of open RyRs $N_{\rm O}^{\rm peak}$ and spark duration $\tau$, remain largely unchanged. This is
expected due to the similarity between the two fitting approaches. When
the value of $K$ in the sigmoid form of the opening rate is halved, the peak number
of open RyRs $N_{\rm O}^{\rm peak}$ remains nearly constant, while some calcium sparks exhibit longer
durations. This aligns with the open probability results in Fig.~\ref{Fig: rate}(c), where a reduction in $K$ increases the open probability of RyRs.

The results for sheep, using the transition rates fitted by Cannell et al., are
shown in Fig.~\ref{Fig: plotrate}(d). These results indicate a higher
occurrence of calcium sparks with longer durations, which is attributed to the
larger opening rates and smaller closing rates for RyRs in sheep, particularly the
continuous decrease in the closing rate as calcium concentration increases. When
we use the sigmoid form for the transition rates, the number of calcium sparks  with longer duration
decreases significantly, as shown in Fig.~\ref{Fig: plotrate}(e). Further
increasing the closing rate from 810.0 s$^{-1}$ to 1500 s$^{-1}$ leads to an
additional reduction in calcium sparks  with longer duration, bringing the results closer to those
seen in rats.

\begin{figure}[htbp!] \includegraphics[bb=0 0 800
  250,scale=0.3,clip]{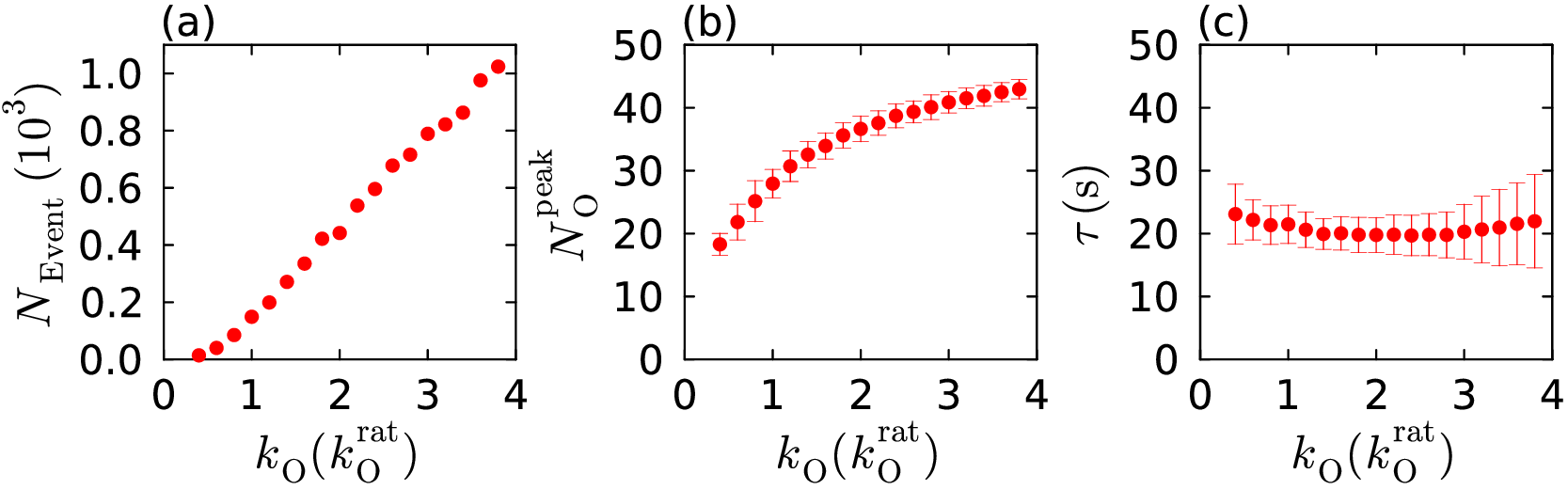}
  \caption{Number of calcium sparks (a), peak number of open RyR (b), and duration (c)  of calcium sparks in clusters of size 50, with varying RyR opening rates, expressed as multiples of the experimentally measured opening rate for rat RyRs. } \label{Fig: open} 
\end{figure}

To further understand the regulation of calcium sparks through variations in the
RyR opening rate, we present in Fig.~\ref{Fig: open} the results of spontaneous
calcium sparks with different RyR opening rates, normalized to the opening rate of
rats. The results indicate that the number of calcium sparks increases rapidly
with higher opening rates. Similarly, the peak number of open RyRs also rises
quickly as the opening rate increases, although this increase slows at higher open
rates. Surprisingly, the effect on spark duration is minimal; after a slight
decrease, the duration stabilizes at around 20~ms. These findings suggest that
increasing the RyR opening rate enhances calcium release through calcium sparks by
increasing both the number of calcium sparks and the number of RyRs activated
within a cluster, while having little impact on spark duration.

With the results presented in Fig.~\ref{Fig: open}, at the experimental opening
rate, approximately 100 spontaneous calcium sparks occur for every 10,000
openings of RyR channels. Considering the estimated number of RyRs in a cell,
which is on the order of $10^6$ as reported by Cheng et al.,~\cite{Cheng1993} if the probability
of spontaneous openings of RyR is 1\%, the experimental observations of
approximately 100 calcium sparks can indeed be achieved.~\cite{Cheng1993}

\subsection{Regulation by calcium concentrations in NSR and myoplasm}

In the current study, we investigate calcium release with calcium concentrations
in the JSR and myoplasm clamped. However, in vivo environments, these calcium
concentrations will vary due to calcium release or other reasons. Now, we
discuss the effect of these calcium concentrations on calcium release. In
Fig.~\ref{Fig: clacium}, we present the distribution of $N_{\rm O}^{\rm peak}$
and $\tau$ with different calcium concentrations in the JSR and myoplasm. In the
simulation, 200 partial distributions of a single cluster with 50 RyRs are
generated, and 10 triggers are performed with time evolution lasting 100~ms (all
CREs with $\tau$ larger than 100~ms will be categorized to 100~ms).

\begin{figure}[htbp!] \includegraphics[bb=15 7 789
750,scale=0.32,clip]{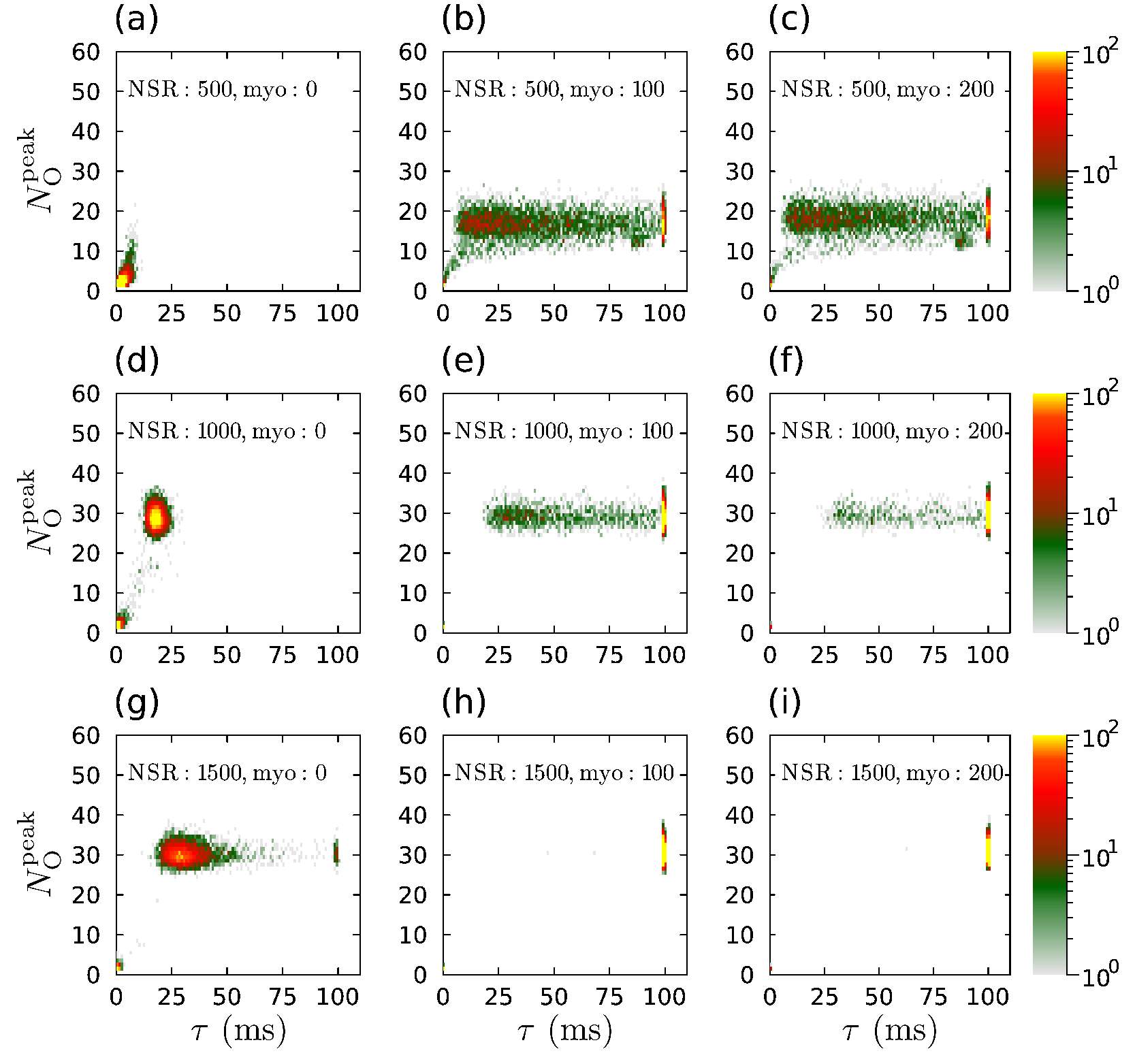} \caption{2D histogram representing the distribution
of CREs on clusters with size 50 against the number of
peak open RyRs $N_{\rm O}^{\rm peak}$ and duration $\tau$. 
The distributions are shown for different levels of calcium concentrations: 
500~$\mu$M (a, b, c), 1000~$\mu$M (d, e, f), and 1500~$\mu$M (g, h, i) in NSR, 
and 0~$\mu$M (a, d, g), 100~$\mu$M (b, e, h), and 200~$\mu$M (c, f, i) in the myoplasm.  The bin sizes
for $N_{\rm O}^{\rm peak}$ and $\tau$ are 1 and 1~ms, respectively.} \label{Fig:
clacium} \end{figure}

For the low calcium concentration in myoplasm, chosen as 0~$\mu$M in simulation,
calcium sparks are difficult to evoke if $[{\rm Ca}^{2+}]^{\rm NSR}$ is set as
500~$\mu$M. As shown in Fig.~\ref{Fig: clacium}(a), most of the CREs have $N_{\rm
O}^{\rm peak}$ smaller than 10 and $\tau$ smaller than 10. With the increase of
$[{\rm Ca}^{2+}]^{\rm NSR}$, larger calcium sparks are evoked. The $N_{\rm
O}^{\rm peak}$ is kept at about 30 and changes little with the $[{\rm
Ca}^{2+}]^{\rm NSR}$ ranging from 1000~$\mu$M to 1500~$\mu$M, while the duration $\tau$
lasts longer. With high $[{\rm Ca}^{2+}]^{\rm NSR}$, the calcium sparks with
long lifetimes become more frequent, and some of them even have a duration of
more than 100~ms with $[{\rm Ca}^{2+}]^{\rm NSR}$ of 1500~$\mu$M. The high calcium
concentration in NSR may lead to a high calcium concentration in JSR $[{\rm Ca}^{2+}]^{\rm JSR}$, resulting
in larger calcium releases from the JSR to SS, which could be the reason for the
longer duration.

For larger calcium concentrations in myoplasm, $[{\rm Ca}^{2+}]^{\rm myo}$ of
100~$\mu$M and 200~$\mu$M, which will lead to a high calcium concentration in SS $[{\rm Ca}^{2+}]^{\rm SS}$. The
sensitivity of the RyR is dependent on $[{\rm Ca}^{2+}]^{\rm SS}$; hence,
calcium release is easier to occur and sustain. For $[{\rm Ca}^{2+}]^{\rm NSR}$
set as 500~$\mu$M, the calcium sparks occur with smaller amplitudes, $N_{\rm O}^{\rm
peak}$ about 20, and the duration has a sparse distribution, ranging from
smaller than 10 ms to larger than 100 ms. With the increase of $[{\rm
Ca}^{2+}]^{\rm NSR}$, the amplitudes become larger and stabilize around 30. The
durations become longer, and most of them are larger than 100 ms at $[{\rm
Ca}^{2+}]^{\rm NSR}$ of 1500~$\mu$M.

\begin{figure}[htbp!] \includegraphics[bb=10 0 800 500,scale=0.3105,clip]{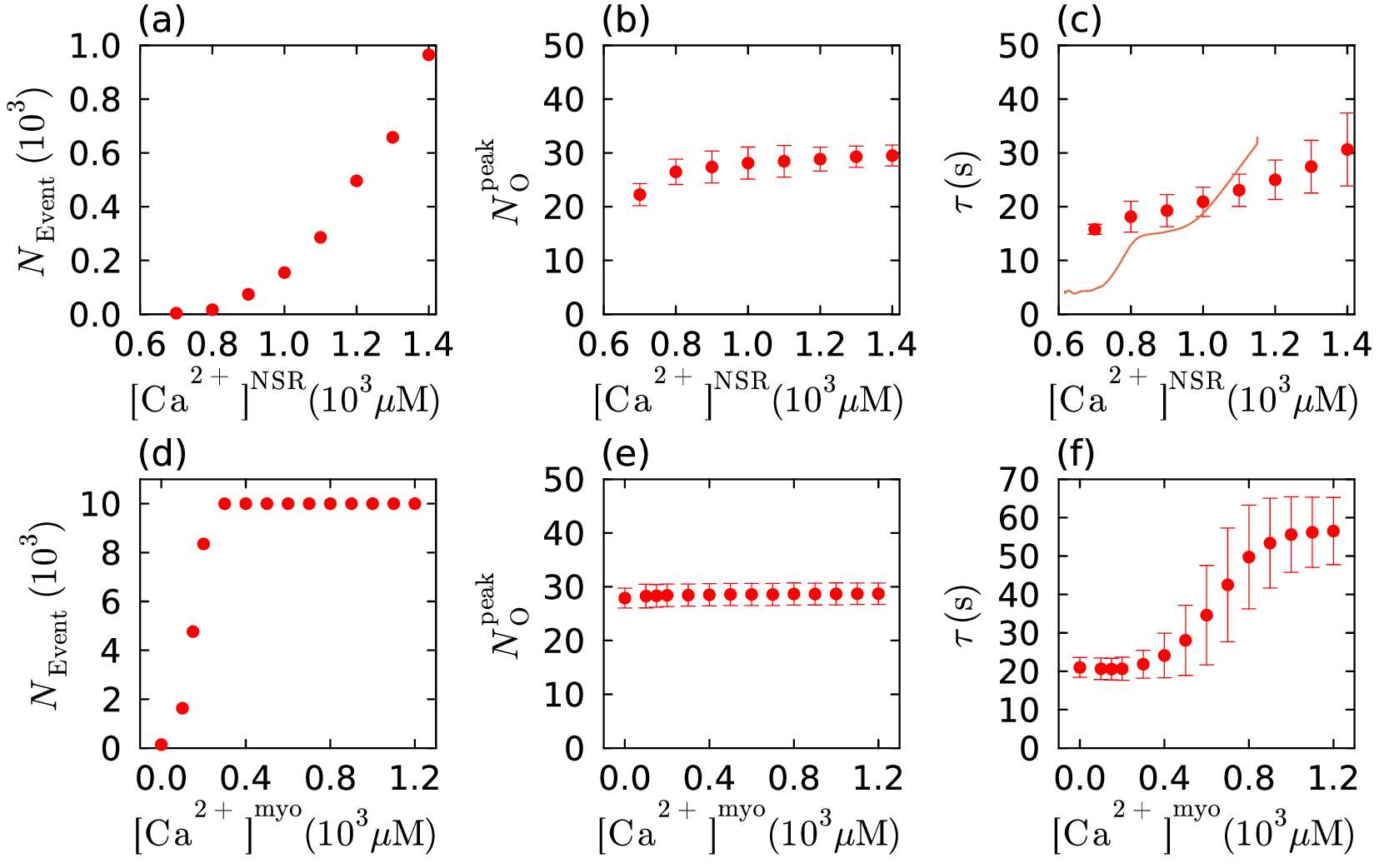}
  \caption{ Number of calcium sparks (a, d), peak number of open RyRs (b, e),
  and duration (c, f) of calcium sparks in clusters of size 50, with varying
  calcium concentrations in the NSR (a, b, c) and myoplasm (d, e, f). The line in (c) is restuls
  by Sato et al.~\cite{Sato2016}
  }
  \label{Fig: plotNSRmyo} 
\end{figure}

To further understand the regulation of calcium sparks under varying calcium
concentrations in the NSR and myoplasm, we present the results of spontaneous
calcium sparks under different calcium levels in Fig.~\ref{Fig: plotNSRmyo}. As
the calcium concentration in the NSR increases, the frequency of spontaneous
calcium sparks rises, as shown in Fig.~\ref{Fig: plotNSRmyo}(a). However, the
peak number of open RyRs $N_{\rm O}^{\rm peak}$ remains mostly stable, with only a slight increase at
lower concentrations, while the duration time decreases gradually. In
Fig.~\ref{Fig: plotNSRmyo}(c), we also provide results of Sato et al.,~\cite{Sato2016}
showing consistency between our findings and theirs, both in terms of the
duration times and the trend as $[{\rm Ca}^{2+}]^{\rm NSR}$ increases.
Additionally, the results in Fig.~\ref{Fig: plotNSRmyo}(a-c) suggest a rapid
increase in flux rate at $[{\rm Ca}^{2+}]^{\rm NSR}$ levels around 600~$\mu$M,
which aligns with the findings of Sato et al.~\cite{Sato2011}

Increasing the calcium concentration in the myoplasm causes a rapid
rise in the number of calcium sparks, reaching 10,000, which corresponds to the
total number of triggers. This indicates that nearly all spontaneous openings of
RyRs trigger calcium sparks. Despite this, the increase in myoplasmic calcium
concentration has little effect on the peak number of open RyRs $N_{\rm O}^{\rm peak}$, which
stabilizes around 30. The duration of calcium sparks remains stable at lower
concentrations but increases sharply with higher concentrations. Though the
duration appears to stabilize again, as shown in Fig.~\ref{Fig: plotNSRmyo}(f),
this is mainly due to the calculation being limited to 60 ms.

These findings suggest that increasing calcium concentrations in both the NSR
and myoplasm primarily enhances calcium release through calcium sparks by
increasing the number of sparks and their duration, with minimal effect on the
number of RyRs activated within a cluster.

\subsection{Regulation  by  Calsequestrin}

The CSQ serves as a critical regulator of calcium release. Its
binding and unbinding from the RyR complex influence the sensitivity of RyRs to
calcium. Additionally, CSQ can bind calcium ions in the JSR, thereby regulating
the calcium concentration within this compartment. In our model, we introduce
binding and unbinding states to simulate the fluctuating sensitivity of RyRs to
CSQ as shown in Fig.~\ref{Fig: state}, while a $\beta$ factor represents the 
buffering effect of CSQ in Eq.~(\ref{Eq: master}) and Eq.~(\ref{Eq: beta}). 

\begin{figure}[htbp!] \includegraphics[bb=7 7 589 500,scale=0.4,clip]{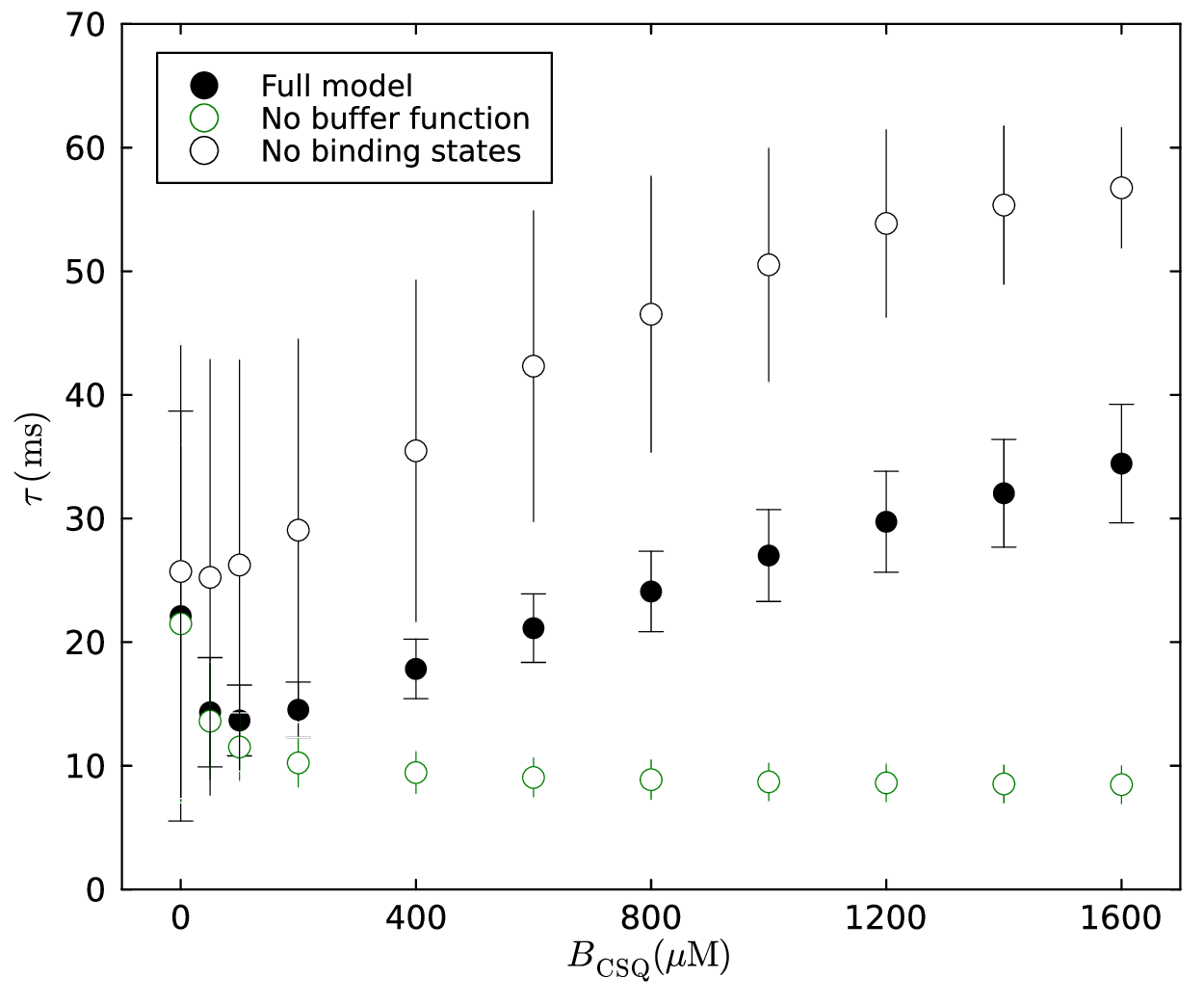} \caption{Variation of the duration of calcium
sparks with CSQ concentration $B_{\rm CSQ}$. The black solid circles represent
the results from the full model, while the green open circles and black open
circles represent models without the buffer function of CSQ and without the two
CSQ binding states, respectively.} \label{Fig: CSQ} \end{figure}

\begin{figure*}[htbp!]
  \includegraphics[bb=0 0 1550 800,scale=0.338,clip]{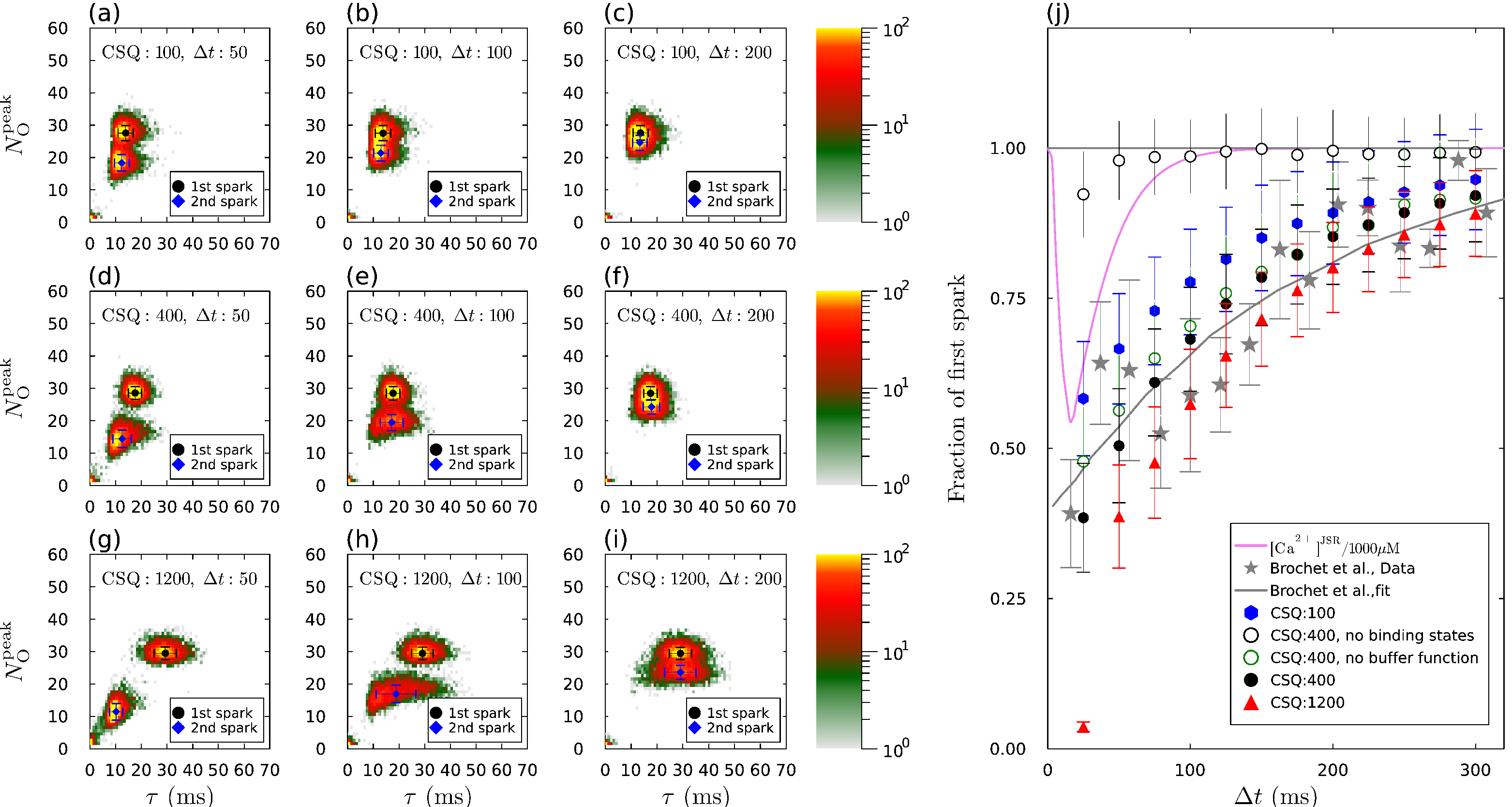}
  \caption{Refractoriness of the calcium sparks with different interval times
  $\Delta t$. (a-i) 2D histogram depicting the distribution of $N_{\rm O}^{\rm
  peak}$ and $\tau$ with different CSQ concentrations and interval times $\Delta t$. 
  The bin sizes for $N_{\rm O}^{\rm peak}$ and $\tau$ are 1
  and 1 ms, respectively. The black circle and blue diamond represent the
  average and uncertainties of $N_{\rm O}^{\rm peak}$ and $\tau$ for the first
  and second sparks. (j) The $N_{\rm O}^{\rm peak}$ of the second spark as the
  fraction of the first spark. The blue hexagon, black full (black empty, green
  empty) circle, and red triangle represent the CSQ concentrations of 100, 400
  (without binding state, without buffer function), 1200~$\mu$M. The solid
  curves depict a typical normalized JSR depletion curve, $[\rm Ca^{2+}]^{\rm
  SS}/100~\mu$M, showing that the recovery from refractoriness is not associated
  with local JSR refilling. Experimental results and fitting (grey star and
  curve) from Brochet et al.~\cite{Brochet2005} }
  \label{Fig: refractoriness}
\end{figure*}

Fig.~\ref{Fig: CSQ} illustrates the variation of durations of CREs corresponding
to changes in CSQ concentration.  The results obtained with a CSQ concentration
of 0~$\mu$M exhibit a widely spread distribution of calcium release events
across the duration $\tau$. As the CSQ concentration increases, this
distribution narrows, indicating a more stable duration for clusters of certain
size. Increasing CSQ levels prolong the calcium release duration from
approximately 10 ms with 100~$\mu$M to 30~ms with 1600~$\mu$M. CSQ plays a dual
role in regulating calcium release: affecting the sensitivity of RyRs to calcium
and serving as a buffer to sequester calcium in the JSR. To elucidate the origin
of the duration variation, we also provide results without the $\beta$ factor in
Eq.~(\ref{Eq: master}), which removes the buffer function, as the green open
circles in Fig.~\ref{Fig: CSQ}.  In this case, the duration remains largely
unchanged with increasing CSQ concentration.  However, if we maintain the buffer
function but remove the binding state, the prolongation of duration can still be
observed as shown as black open circles in Fig.~\ref{Fig: CSQ}. This verifies that
the variation in duration is primarily due to the buffering action of CSQ inside
the SR, as suggested by Terentyev et al.~\cite{Terentyev2003}

As depicted in Fig.~\ref{Fig: Single0}, a calcium spark lasts approximately 20~ms,
after which the RyRs close and the calcium concentration in the SS returns to
its resting state. However, the calcium blink in the JSR persists longer after
the RyR closure and termination of the calcium sparks in the SS. During this
period, the calcium concentration in the JSR $[{\rm Ca}^{2+}]^{\rm JSR}$ does
not fully recover to 1000~$\mu$M, reducing the likelihood of subsequent calcium
sparks. As simulated in our study and by other researchers, the calcium blink
typically occurs on a timescale of about 100~ms. Interestingly, experimental
measurements of refractoriness are often much longer than those derived from the
calcium blink.~\cite{Brochet2005} Such excess refractoriness may result from the
CSQ affecting the activity of RyRs. In our model, the regulation of RyRs by CSQ
is accomplished through two binding states, as described by Restrepo et al.~\cite{Restrepo2008} Here, we present our findings regarding refractoriness
in Fig.~\ref{Fig: refractoriness}.

In the simulation, the RyRs are activated by LCC calcium release in model II scenario,
wherein a cluster of 50 RyRs is paired with 50/7 LCCs. Such activations ensure
that a calcium spark can be evoked with a trigger. After an interval $\Delta t$ following
the termination of the first calcium spark, another activation is performed to
trigger the second calcium spark. We generate 500 partial distributions of the
cluster, and for each distribution, 10 simulations are performed. The $N_{\rm
O}^{\rm peak}$ and $\tau$ of calcium sparks are collected.

In Fig.~\ref{Fig: refractoriness}(a-i), two groups of CREs are evident. The group with larger $N_{\rm
O}^{\rm peak}$ corresponds to the first sparks, while the other with smaller
$N_{\rm O}^{\rm peak}$ represents the second sparks. Comparing the subfigures
for the same CSQ concentration, one can observe that with increasing intervals,
the two groups gradually merge into one, indicating the disappearance of
refractoriness. Moreover, as the CSQ concentration increases, the deviation of
the second spark from the first sparks becomes larger.

The $N_{\rm O}^{\rm peak}$ of the second spark as the fraction of the first
spark is shown in Fig.~\ref{Fig: refractoriness}(j) and compared with changes in
calcium concentration in the JSR as well as experimental results. This figure
illustrates that at high levels of CSQ concentration, the fraction of first
sparks resulting in second sparks becomes smaller, consistent with findings by
Restrepo et al.~\cite{Restrepo2008} Our model reproduces these observed
properties, aligning with the experimental findings of Terentyev et
al.,~\cite{Terentyev2003} which demonstrate that the refractory period increases
when CSQ is overexpressed in transgenic mice.  To confirm that the increased
spark refractoriness is not caused by the refilling of local JSR stores, we plot
a typical JSR depletion curve, i.e., $[\rm Ca^{2+}]^{\rm SS}/100~\mu$M, as a
function of time. As shown in the figure, the JSR stores refill much faster than
the time it takes for the second spark to reach the same level as the first
spark. This indicates that factors beyond simple ion replenishment are
contributing to the refractory period.

To investigate the origins of the excess refractoriness, we also provide results
with the removal of the $\beta$ factor and the binding state. These results
differ significantly from the variation of duration with CSQ concentration
described above. Without the buffer function, the fraction remains almost the
same as in the full model. However, if we retain the buffer function but remove
the binding state, the excess refractoriness vanishes, and the fractions
increase to 1 as $[\rm Ca^{2+}]^{\rm SS}/100~\mu$M.  These findings suggest that
the excess refractoriness is primarily due to the CSQ-dependent sensitivity of
the RyR, rather than the buffering action of this protein inside the JSR. The
CSQ binding state appears to play a crucial role in modulating RyR activity,
leading to prolonged refractory periods. When the CSQ binding state is removed,
RyRs regain their normal sensitivity more quickly, resulting in a faster
recovery and a higher likelihood of subsequent sparks occurring at a similar
magnitude to the initial spark.

\subsection{CREs under malfunctioning calsequestrin regulation}

\begin{figure}[htbp!]  
  \includegraphics[bb=0 7 789 450,scale=0.31,clip]{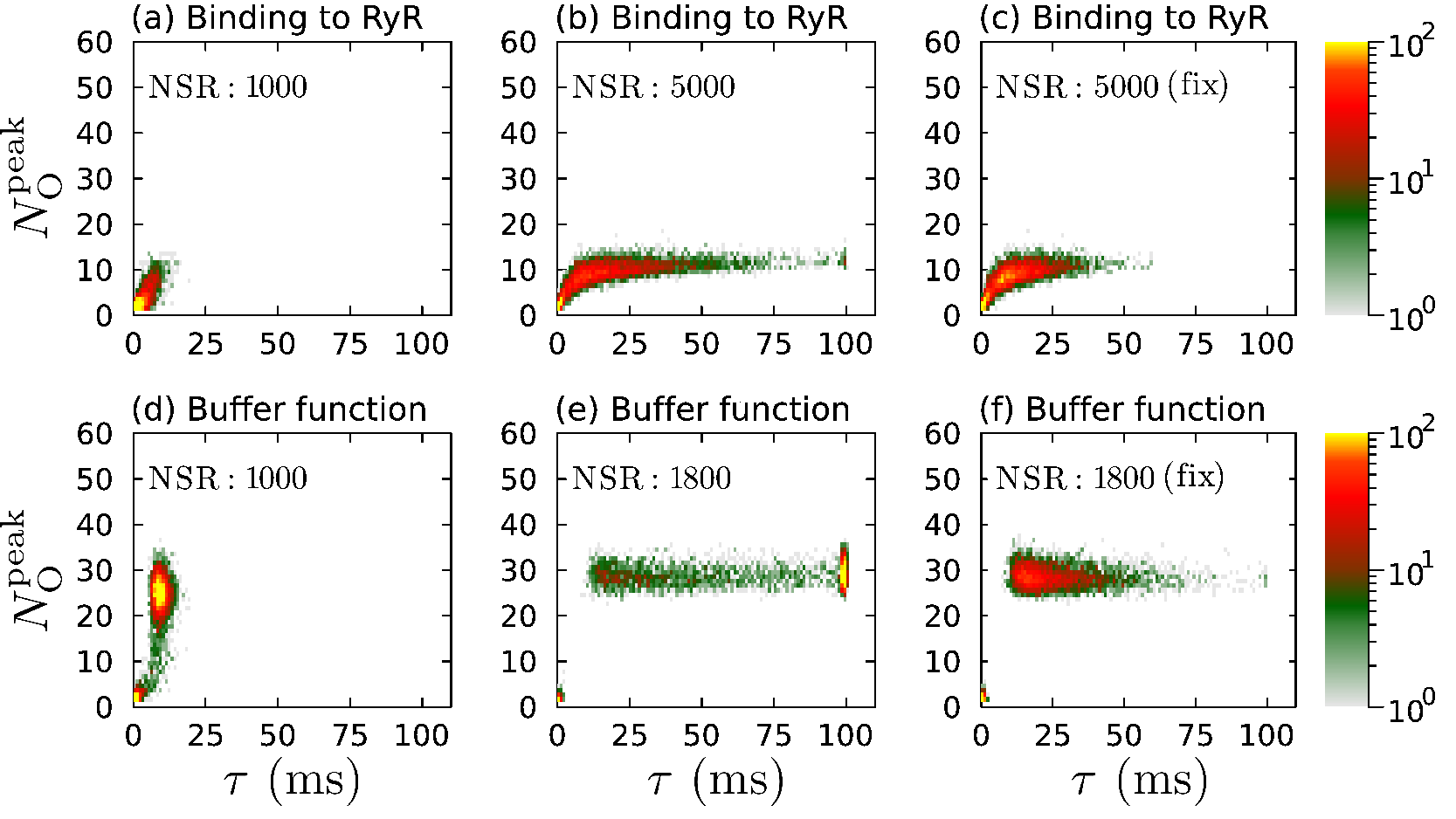}  
  
  \caption{2D histogram representing the distribution of CREs under CSQ
  malfunction conditions. Panels (a-c) illustrate the scenario where all RyR are
  bound by CSQ, while panels (d-f) depict a malfunction in the buffering
  function of CSQ. Panels (a,d), (b,e), and (c,f) correspond to $[\rm
  Ca^{2+}]^{\rm NSR}$ concentrations of 1000 and 5000 (1800)~$\mu$M for different
  spatial distributions of RyR clusters, and 5000 (1800)~$\mu$M for a fixed spatial
  distribution, respectively.}  
  
  \label{Fig: malfunction}  
\end{figure}

The simulations presented above suggest that the CSQ has a dual role in
regulating calcium release. If the CSQ regulation mechanism malfunctions, the
calcium release becomes irregular. Here, we investigate two malfunctioning
scenarios that result in a decrease in calcium release: CSQ release dysfunction
from the RyR complex, as shown in Fig.~\ref{Fig: malfunction}(a-c), and malfunction
of the buffering function of CSQ, as depicted in Fig.~\ref{Fig: malfunction}(d-f).
During the simulations, 500 partial distributions of a cluster comprising 50
RyRs were generated, and 10 triggers were applied with a time evolution of 100
ms (all CREs with $\tau$ values greater than 100 ms were categorized as having a
duration of 100 ms).

In Fig.~\ref{Fig: malfunction}(a-d), the calcium concentrations in the NSR and
myoplasm are set to standard values of 1000 and 0.1$\mu$M, respectively. When
the RyR complex is bound with CSQ, it exhibits low activity and is less likely
to open. If the CSQ malfunctions when releasing from the RyR complex, the RyR
becomes harder to open, resulting in very short-duration CREs and few open RyRs,
as seen in Fig.~\ref{Fig: malfunction}(a).  On the other hand, if the CSQ cannot
buffer calcium in the JSR, a significant number of RyRs within a cluster still
open, but with a short duration, as shown in Fig.~\ref{Fig: malfunction}(b).
Both of these malfunctions in CSQ regulation reduce calcium release from the
JSR.  To increase calcium release, the $[\rm Ca^{2+}]^{\rm NSR}$ was increased
to 5000 and 1800~$\mu$M in two scenarios, respectively. This indeed enhanced
calcium release in both scenarios [Fig.~\ref{Fig: malfunction}(b,e)].  However,
one can observe that the distribution of CRE durations becomes more dispersed,
indicating that the on-off switch characteristics of cluster cannot be fully
recovered to those observed with standard $[\rm Ca^{2+}]^{\rm NSR}$ value. To
further investigate whether this dispersion is due to differences in spatial
distribution, we fixed the spatial distribution of the RyR cluster and performed
5000 triggers. As seen in Fig.~\ref{Fig: malfunction}(c, f), the results still
exhibit dispersion, indicating that the on-off switching characteristic of a RyR
cluster is not fully restored for a RyR cluster, even with increased calcium release.

\section{Discussion and summary}\label{Summary}

The RyRs form clusters that exhibit irregular shapes, varying sizes, and a
stochastic distribution of receptors. In the current study, we have depicted the
spatial distribution of RyRs using a nonlinear spatial network. By integrating
calcium exchange mechanisms and calcium buffering dynamics, we have successfully
constructed a model that simulates calcium release through RyRs located on the
JSR under conditions of clamped calcium concentrations in the NSR and myoplasm.

\subsection{RyR Cluster On-Off Behavior}

Our model successfully reproduces the characteristic on-off behavior of RyR
clusters during calcium release. In spontaneous simulation scenarios, RyRs open
briefly for approximately 0.001ms, with only a few remaining open for over 1ms,
potentially triggering CREs. These CREs can be
classified into two categories: short-lived calcium quarks, lasting less than
10ms with only a few activated RyRs, and longer-lasting calcium sparks,
involving significantly more RyRs. These findings confirm that RyR clusters
operate as on-off switches in calcium release, as suggested by prior
studies.\cite{Asfaw2013,Song2016} While the probability of initiating a calcium
spark from random RyR openings is low, our simulations show that sustained
LCC release for 6.7ms can reliably trigger calcium
sparks. These sparks exhibit similar characteristics to spontaneous sparks in
terms of duration and other key properties, consistent with experimental
observations.\cite{Cheng1993,Wang2001,Lopez-Lopez1995} Notably, this behavior is
robust across various LCC triggers, further validating the model's predictive
capabilities.

\textbf{Mechanisms Governing Spark Termination:} The duration of calcium sparks
is determined primarily by their termination mechanisms, which have been widely
studied in the literature. Proposed explanations include local calcium depletion
in JSR and "stochastic
attrition".~\cite{Cheng1996,Gyoke1998,Stern1992} In our model, spark termination
is governed by the reduction of the calcium gradient between the subspace and
JSR, as well as the stochastic opening and closing of RyRs. This suggests that
termination is not solely dependent on complete JSR
depletion or stochastic attrition.  During a calcium spark, the calcium
concentration in the JSR decreases but stabilizes at approximately 500~$\mu$M.
Interestingly, when the JSR calcium concentration falls below 500~$\mu$M,
CSQ begins to bind to RyRs, reducing RyR activity and
facilitating spark termination. This mechanism highlights the role of CSQ in
modulating calcium release dynamics.

\textbf{Modulation by Species-Specific RyR Properties:} Cannell et al. revealed
notable differences in RyR behavior between rats and sheep, particularly in the
transition rates affecting CREs.~\cite{Cannell2013} For rat RyRs, both the
transition rates used in the original study and a newly proposed sigmoid fit
show a concentration of CREs. However, reducing the RyR opening rate parameter
can extend spark durations in certain cases.  In contrast, for sheep RyRs,
the transition rates suggested by Cannell et al.~\cite{Cannell2013} result in
longer calcium sparks due to a continuous decrease in the closing rate with
increasing calcium concentration, which appears unrealistic. By adjusting to a
constant closing rate in the new sigmoid fit, the results align more closely
with observations in rats. Furthermore, variations in RyR opening rates increase
the number of calcium sparks and activated RyRs with minimal impact on spark
duration, suggesting that increased RyR opening primarily facilitates calcium
release.

\textbf{Calcium Concentration Effects on CRE Dynamics:} The resting myoplasmic
calcium concentration is approximately 0.1~$\mu$M, which is critical for
maintaining stable calcium sparks with a duration of around 20~ms. When the
myoplasmic calcium concentration is increased to 100~$\mu$M, the distribution of
spark durations becomes sparse, with some lasting over 100~ms.  Higher calcium
concentrations in both the myoplasm and the NSR are necessary to
effectively reproduce calcium sparks. Lower NSR calcium concentrations result in
CREs with short durations or sparse distributions, while higher NSR calcium
concentrations lead to significantly prolonged CREs. Additionally, increased
calcium release elevates myoplasmic calcium levels, further prolonging CRE
durations and potentially triggering intracellular calcium oscillations.  These
findings suggest that elevating calcium concentrations in the NSR and myoplasm
enhances calcium release by increasing the number and duration of calcium
sparks, with minimal impact on the number of RyRs activated within a cluster.

\subsection{CSQ as a Key Modulator of  Calcium Release}

The intricate role of CSQ in calcium release regulation is
critical for understanding the pathophysiology of various
diseases.~\cite{Liu2009, Sibbles2022, Sun2021} This study investigates two key
mechanisms through which CSQ influences calcium dynamics: its role as a calcium
buffer within the JSR and its interaction with RyRs to modulate their activity.
Using computational simulations, we examined the effects of varying CSQ
concentrations on CRE durations and the phenomenon of excess refractoriness.

\textbf{CSQ as a Buffer:} Our results confirm the dual role of CSQ in calcium
regulation, consistent with prior studies.\cite{Damiani1994, Chen2013, Sun2021}
Specifically, variations in CSQ concentrations significantly influence CRE
durations, in agreement with experimental observations.\cite{Terentyev2003} By
selectively removing CSQ's buffering function and its binding interactions, we
found that its primary influence on CRE durations arises from its buffering
capacity in the JSR. This underscores the pivotal role of CSQ in shaping the
temporal profile of calcium release by maintaining calcium homeostasis within
the JSR.

\textbf{CSQ-RyR Interaction:} In addition to its buffering role, CSQ interacts
with RyRs to regulate their activity and sensitivity, a mechanism closely tied
to excess refractoriness observed experimentally.~\cite{Brochet2005} Our
simulations demonstrate that removing CSQ-RyR binding states led to a
significant reduction in refractory periods. This indicates that the interaction
between CSQ and RyRs is critical for modulating RyR sensitivity and regulating
calcium release dynamics. These findings highlight the complex interplay between
CSQ and RyRs in stabilizing calcium signaling within cardiomyocytes.

\textbf{Impacts of CSQ Dysregulation:} Further analysis revealed the impact of
CSQ dysregulation on calcium release dynamics. Excessive binding of CSQ to RyRs,
coupled with the loss of its buffering function, impeded the proper opening of
RyRs. While an increase in NSR calcium concentrations partially restored calcium
release, the durations of calcium sparks exhibited a broad range of variability.
More importantly, the characteristic on-off switching behavior of RyR clusters
was not restored under these conditions. This indicates that CSQ dysregulation
disrupts calcium release dynamics and undermines the functional integrity of RyR
clusters.

\subsection{Conclusions and Future Perspectives}

This study highlights the critical role of RyR clusters in calcium release, emphasizing their robust on-off switch behavior. This switching mechanism forms the foundation of calcium dynamics, with CSQ serving as a key modulator that influences the temporal and spatial characteristics of calcium release through its buffering capacity and interaction with RyRs. Dysregulation of CSQ functions leads to significant disruptions in calcium signaling, shedding light on its potential involvement in the pathophysiology of cardiac diseases.

Our findings demonstrate that, under calcium-clamped conditions with a realistic RyR2 network, RyR clusters exhibit robust spark structures of varying sizes and achieve switch-like behavior without requiring inactivation mechanisms. These results underscore the importance of network realism in facilitating the emergence of switch-type behavior in RyR clusters.

While this study provides valuable insights, it is important to acknowledge that the calcium-clamped setup represents a simplified system. Future research should extend these findings by exploring RyR dynamics under more physiologically relevant cycling conditions, where calcium concentrations fluctuate continuously. Investigating the interplay between RyR clusters and CSQ in such dynamic scenarios will be crucial for fully understanding calcium signaling mechanisms in cardiomyocytes.

Moreover, future studies should focus on whether the robust switch-type behavior observed here can be sustained under real cycling conditions, and on the specific roles CSQ plays in modulating RyR cluster activity in such systems. By addressing these questions, we can deepen our understanding of RyR-mediated calcium release and develop more comprehensive models of calcium signaling in health and disease.

\begin{acknowledgments}
This project is supported by the National Natural Science
Foundation of China (Grant No. 12475080,11675228).
\end{acknowledgments}

\section*{Data Availability Statement}

The data that support the findings of this study are available within the article as well as from the corresponding author upon reasonable request.

%\nocite{*}
\bibliography{aipsamp}% Produces the bibliography via BibTeX.
\end{document}